\documentclass[12pt,a4paper]{article}
\usepackage{amsmath}
\usepackage{graphicx,amsmath,amssymb}

\usepackage{cite}
\usepackage{graphicx,amsmath,amssymb}

\DeclareMathAlphabet{\mathpzc}{OT1}{pzc}{m}{it}

\newcommand{\K}{\mathcal{K}}

\newcommand{\be}{\begin{equation}}
\newcommand{\ee}{\end{equation}}
\newcommand{\bi}{\begin{itemize}}
\newcommand{\ei}{\end{itemize}}

\date{}
{ 
\title{ \bf  \Large  Description of $F_2$ and $F_L$ at small $x$ using a collinearly-improved BFKL resummation} 

\author{Martin Hentschinski$^{1}$, Agust{\' \i}n Sabio Vera$^{2}$, Clara Salas$^{2}$ 
\bigskip \\  \normalsize 
 $^1$Physics Department,  Brookhaven National Laboratory,  \\   \normalsize  Upton, NY 11973, USA\\ \normalsize
$^2$ Instituto de F{\' \i}sica Te{\' o}rica UAM/CSIC 
and   \\  \normalsize   % \\ \it
 Universidad Aut{\' o}noma de Madrid,
E-28049 Madrid, Spain
}

}
\begin{document} 

\maketitle

\begin{abstract}
We present a detailed description of the $Q^2$ and $x$ dependence of the 
structure functions $F_2$ and $F_L$ as extracted from the Deep Inelastic Scattering data at HERA in 
the small Bjorken $x$ region. Making use of a collinearly-improved BFKL equation at next-to-leading 
order and a treatment of the running of the coupling using non-Abelian physical renormalization 
together with the BLM scale choice allows us to reach low values of $Q^2$. We also provide some 
predictions for future lepton-hadron colliders. 
\end{abstract}

\section{Introduction \& review of our framework}

Recently, in~\cite{Hentschinski:2012kr}, we calculated the effective
slope $\lambda$ for the structure function $F_2$ when parameterized as
$\sim x^{-\lambda}$ at low Bjorken $x$, which ranges from $\lambda
\simeq 0.1$ to $\simeq 0.3$ when moving from low to high values of
$Q^2$. The importance of a correct understanding of the HERA data
using perturbative QCD techniques, in particular for the physics
program at the Large Hadron Collider (LHC), has been largely discussed
in the literature (see, {\it e.g.}, Ref.~\cite{reviews}).

In this Letter we will not only further investigate the $Q^2$ and $x$
dependence of $F_2$ but also study in detail the longitudinal
structure function $F_L$. As in~\cite{Hentschinski:2012kr} we will
explore the small $x$ region using the next-to-leading order
(NLO)~\cite{Fadin:1998py} BFKL~\cite{BFKL1} equation with collinear
improvements, together with optimal renormalization schemes.

Let us first briefly review our
formul{\ae}~\cite{Hentschinski:2012kr}. At small $x \simeq Q^2/s$, with $s$ being the 
squared center-of-mass energy, we
can apply high energy factorization and write the structure functions
$F_I$, $I = 2, L$ as (note that the integrations take place in the
transverse space with $q \equiv \sqrt{{\bf q}_\perp^2}$)
\begin{eqnarray}
F_I  (x, Q^2)&=&  \int \frac{d^2 {\bf q}_\perp}{ \pi \, q^2} \int \frac{d^2 {\bf p}_\perp}{ \pi \, p^2} 
\Phi_I \left(q,Q^2\right) \Phi_P \left(p,Q^2_0\right) {\cal F} \left(x, q , p\right).
\end{eqnarray}
$\Phi_P $ is the non-perturbative proton impact factor which we model using
\begin{eqnarray}
\label{eq:protoN}
\Phi_P \left(p,Q_0^2\right) &=& \frac{ {\cal C}}{\Gamma( \delta)} \left(\frac{p^2}{Q_0^2}\right)^\delta e^{-\frac{p^2}{Q_0^2}},
\end{eqnarray}
where we have introduced two free parameters and a normalization. $\Phi_I$ is the impact factor associated to the photon which 
in~\cite{Hentschinski:2012kr} we treated at leading-order (LO), {\it i.e.}
\begin{align}
\label{eq:1}
\int \frac{d^2 {\bf q}_\perp}{ \pi q^2}  \Phi_I \left(q,Q^2\right) \left({\frac{q^2}{  Q^2} }\right)^{\gamma-1}  &= \;\; \frac{{\alpha}_s(\mu^2) }{2\pi}
\sum_{q=1}^{n_f} e_q^2 \; c_I(\nu) \; ,
\end{align}
where 
\begin{align}
c_I(\nu) & \equiv \frac{\pi^2}{4} \frac{\Omega_I(\nu)}{(\nu+ \nu^3)}\, {\rm sech}{(\pi \nu)}\, \tanh{(\pi \nu)} 
%\qquad \Omega_L = \nu^2+ 1/4 \qquad \Omega_2 = (11+12 \nu^2)/8\; ,
\end{align}
$\nu = i(1/2 - \gamma)$, $\Omega_2 = (11+12 \nu^2)/8$, $\Omega_L = \nu^2+ 1/4$, and the strong coupling $\alpha_s$ is fixed at the renormalization scale $\mu^2$. In the present work we will also use the kinematically improved impact factors proposed
in~\cite{Kwiecinski:1997ee,Bialas:2001ks}, which include part of the
higher order corrections by considering exact gluon kinematics.  Its
implementation requires to replace the functions $c_I(\nu)$ by  $\tilde{c}_I(\gamma, \omega)$ where
\begin{eqnarray}
  \tilde{c}_L(\gamma,\omega) = \frac{4 \Gamma(\gamma\! +\! \xi\! +\! 1) \Gamma(1\! +\! \gamma)
   \left[ \left(\psi(\gamma\! +\!\xi) - \psi(\gamma)\right)\left(3 \omega^2 - \xi^2 + 1\right) - 6 \omega \xi \right]}{ \xi \Gamma(1 + \omega)(\xi^4 -5 \xi^2 + 4)}
\end{eqnarray}
and $\tilde{c}_2 = \tilde{c}_L + \tilde{c}_T$, with 
\begin{align} 
&\tilde{c}_T(\gamma, \omega) = 
\frac{\Gamma(\gamma\! +\! \xi)\Gamma(\gamma)}{\xi \Gamma(1\! +\! \omega) (\xi^4 - 5 \xi^2 + 4)}
\bigg\{ -2 \xi\omega \left( \xi^2  + 3 ^2 + 6 \omega + 11  \right)   \notag \\
%&&\hspace{-13.5cm} 
& + \big[ \psi(\gamma + \xi) - \psi(\gamma)\big] \big[ \xi^4 - 10 \xi^2 + 3\omega^2 \left( \omega^2 + 2 \omega + 4   \right) - 2 \omega\left( \xi^2\! -\!1 \right) + 9       \big] 
%\left( 3\omega^2 (\omega + 1)^2 + 9 \omega^2  - 2 \omega (\xi^2 -1) +   \xi^4 - 10 \xi^2 + 9   % (\xi^2 -1)(\xi^2 - 9)
%\right)
\hspace{-.1cm}
 \bigg\}. 
\end{align}
$\psi(\gamma)$ is the logarithmic derivative of the Euler Gamma
function and $\xi = 1 - 2 \gamma + \omega$, while $\omega$ is the Mellin
variable conjugate to $x$ in the definition of the gluon Green function
${\cal F}$, see Eq.~\eqref{eq:2} below. The main difference between
these impact factors is that the LO ones roughly double the value of
their kinematically improved counterparts in the region with small
$|\nu|$, while being very similar for $|\nu| \geq 1$.

The gluon Green function can be written in the form
\begin{align}
\label{eq:2}
{\cal F} \left(x, q , p\right) &=\frac{1}{\pi} \int  \frac{d \omega}{2 \pi i} \int \frac{d \gamma}{2 \pi i} \quad
\frac{1}{ q^2}  \left(\frac{{q^2}}{p^2}\right)^{\gamma} x^{-\omega}
  \frac{1}{\omega- \bar{\alpha}_s \hat{\K} \left(\gamma\right)} ,
\end{align}
with $\bar{\alpha}_s = \alpha_s N_c/ \pi$. The collinearly improved BFKL kernel as introduced in 
eq.~(\ref{eq:2}) is an operator consisting of a diagonal (scale invariant) piece $\hat{\chi} (\gamma)$ with eigenvalue 
\begin{eqnarray}
\label{eq:3}
\chi(\gamma) &= &  \bar{\alpha}_s \chi_0 (\gamma) +\bar{\alpha}_s^2 \chi_1 (\gamma) 
- \frac{1}{2} \bar{\alpha}_s^2 {\chi_0}'  (\gamma) \chi_0 (\gamma) 
+ \chi_{\text{RG}}(\bar{\alpha}_s, \gamma, a, b) \; ,
\end{eqnarray} 
where  $\chi_0(\gamma) = 2 \psi(1) - \psi(\gamma)-\psi(1-\gamma)$, $a = \frac{5}{12} \frac{\beta_0}{N_c} - \frac{13}{36} \frac{n_f}{N_c^3}- \frac{55}{36}$ and  
$b = - \frac{1}{8} \frac{\beta_0}{N_c} - \frac{n_f}{6N_c^3}- \frac{11}{12}$, plus a term $\hat{\chi}_{\text{RC}} (\gamma)$ proportional to $\beta_0$ which contains the running coupling corrections of the 
NLO kernel~\cite{Vera:2007dr}: 
\begin{eqnarray}
\label{eq:33}
 \hat{\chi}_{\text{RC}} (\gamma) & =& 
 \bar{\alpha}_s^2 \frac{\beta_0}{8N_c}  \left( {\chi}_0 (\gamma) \overrightarrow{\partial}_{\gamma} -\overleftarrow{\partial}_{\gamma} {\chi}_0 (\gamma) 
+ 2 \log( \mu^2) \right).
\end{eqnarray}
 The precise form of the NLO  kernel $\chi_1$ can be found in~\cite{Hentschinski:2012kr}. The resummation of
collinear logarithms of order $\bar{\alpha}^3_s$  and beyond is realized by the term~\cite{Salam:1998tj,Vera:2005jt,Hentschinski:2012kr}
\begin{eqnarray}
\label{eq:RG}
&& \chi_{\text{RG}}(\bar{\alpha}_s, \gamma, a, b)
\,\, =  \,\,\bar{\alpha}_s (1+ a \bar{\alpha}_s) \left(\psi(\gamma) 
- \psi (\gamma-b \bar{\alpha}_s)\right) \nonumber\\
&& \qquad \quad - \frac{\bar{\alpha}_s^2}{2} \psi'' (1-\gamma)  - b \bar{\alpha}_s^2 \frac{\pi^2}{\sin^2{(\pi \gamma)}}
+ \frac{1}{2} \sum_{m=0}^\infty \Bigg(\gamma-1-m+b \bar{\alpha}_s  \nonumber\\
&&
\qquad \quad - \frac{2 \bar{\alpha}_s (1+a \bar{\alpha}_s)}{1-\gamma+m}
+ \sqrt{(\gamma-1-m+b \bar{\alpha}_s)^2+ 4 \bar{\alpha}_s (1+a \bar{\alpha}_s)} \Bigg).
\end{eqnarray}
 Our final  expression for the structure functions reads 
 \begin{eqnarray}
 \hspace{-.3cm}F_I (x,Q^2) \hspace{-.3cm} &\propto & \hspace{-.3cm} \int\!\! d\nu \, x^{-{\chi} \left(\frac{1}{2}+i \nu\right)} \Gamma\left(\delta-\frac{1}{2}-i \nu \right)
\Bigg[1 + \frac{\bar{\alpha}_s^2\beta_0 \chi_0 \left(\frac{1}{2}+i \nu\right) }{8 N_c} \log{\left(\frac{1}{x}\right)}  \\
&& \hspace{-2.3cm}
\times \left(i (\pi {\rm coth} (\pi \nu)-2 \pi \tanh{(\pi \nu)}- M_I (\nu))- \psi \left(\delta-\frac{1}{2} - i \nu\right)\right)\Bigg] \left(\frac{Q^2}{Q_0^2}\right)^{\frac{1}{2}+i \nu}\hspace{-.7cm} c_I (\nu),\nonumber
\label{Frho}
\end{eqnarray}
where $M_2$ and $M_L$ can be found in~\cite{Hentschinski:2012kr}. For the kinematical improved version of $F_I$ we replace $c_I (\nu)$ by $\tilde{c}_I (1/2+i\nu,\chi (1/2+i\nu))$.
%\begin{eqnarray}
%\tilde{c}_2 (\omega, \gamma) &=& \frac{2}{\pi} \left( H_L (\gamma,\omega, 1-2\gamma+\omega)+ H_T (\gamma,\omega, 1-2\gamma+\omega)\right) \;\;\; \text{and} \nonumber \\ 
%\tilde{c}_L (\omega, \gamma) &=& \frac{2}{\pi} H_L (\gamma,\omega, 1-2\gamma+\omega) \; .
%\end{eqnarray}
 
 In Eq.~\eqref{Frho}  the scale of the running of the coupling has been set to 
$\mu^2 = Q Q_0$. Building on the work of~\cite{Brodsky:1998kn} we found in~\cite{Hentschinski:2012kr} that in 
order to obtain a good description of the $Q^2$ dependence of the effective intercept of $F_2$, $\lambda$, for  
$x < 10^{-2}$, it is very useful to operate with  non-Abelian physical renormalization schemes using the 
Brodsky-Lepage-Mackenzie (BLM) optimal scale setting~\cite{Brodsky:1982gc} with the momentum space (MOM) physical renormalization scheme~\cite{Celmaster:1979km}. For technical details on our 
precise implementation we refer the reader to~\cite{Hentschinski:2012kr} (see also~\cite{Brodsky:2011ig} for a review on the subject and~\cite{joseagus} for a related work). More qualitatively, in these schemes the pieces of the NLO BFKL kernel proportional to $\beta_0$ are absorbed in a new definition of the running coupling in order to remove the infrared renormalon ambiguity. Once this 
is done, the residual scheme dependence in this framework is very small. We also found it convenient~\cite{Hentschinski:2012kr} to introduce, in order to describe the data with small $Q^2$, an analytic parametrization of the running 
coupling in the infrared proposed in~\cite{Webber:1998um}.

\section{Comparison to DIS experimental data}

In the following we compare our  results with the
experimental data for $F_2$ and $F_L$.
\subsection{$F_2$}

\begin{figure}[htbp]
  \centering
  \vspace{1.cm}
  \includegraphics[scale=.9]{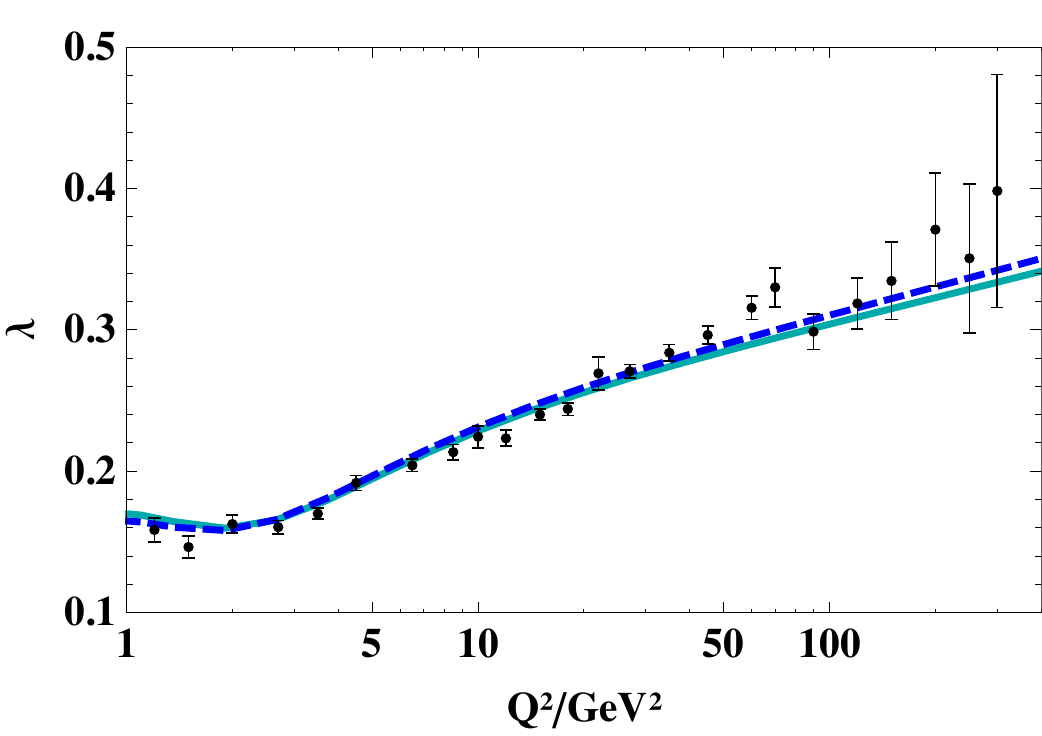}
  \caption{Fit to $\lambda$ for $F_2$ with the LO photon impact factor (solid line) and the kinematically improved one (dashed line). The data set has been extracted from~\cite{Aaron:2009aa}. }
  \label{fig:improved_photon}
\end{figure}

Let us first compare the result obtained in~\cite{Hentschinski:2012kr} for the logarithmic
derivative $d \log F_2 / d \log (1/x)$ using Eq.~(\ref{Frho}) with a LO photon impact factor and 
our new calculation using the kinematically improved one.  In Fig.~\ref{fig:improved_photon} we present our results 
with the values of our best fits for both types of impact factors and compare them with the H1-ZEUS combined data~\cite{Aaron:2009aa} for $x<10^{-2}$. The values of the parameters defining the proton 
impact factor in~(\ref{eq:protoN}) and the position of the (regularized) Landau pole (we use $n_f=4$) for the strong coupling are $\delta = 8.4$, $Q_{0} = 0.28$ GeV, $\Lambda = 0.21$ GeV for the LO order case and $\delta = 6.5$,  $Q_{0} = 0.28$ GeV, $\Lambda = 0.21$ GeV for the kinematically improved (note that the normalization ${\cal C}$ does not contribute to this quantity).  

The LO impact factor generates lower values than the kinematically improved one in the high $Q^2$ region and slightly higher ones when $Q^2 \lesssim 2 \, {\rm GeV}^2$. It is interesting to see how the approach presented here allows for a good description of the data in a very wide 
range of $Q^2$, not only for high values, where the experimental uncertainties are larger, but also in the 
non-perturbative regions due to our treatment of the running of the coupling.

\begin{figure}[htbp]
  \centering {\hspace{-1.5cm}
 \includegraphics[scale=.5]{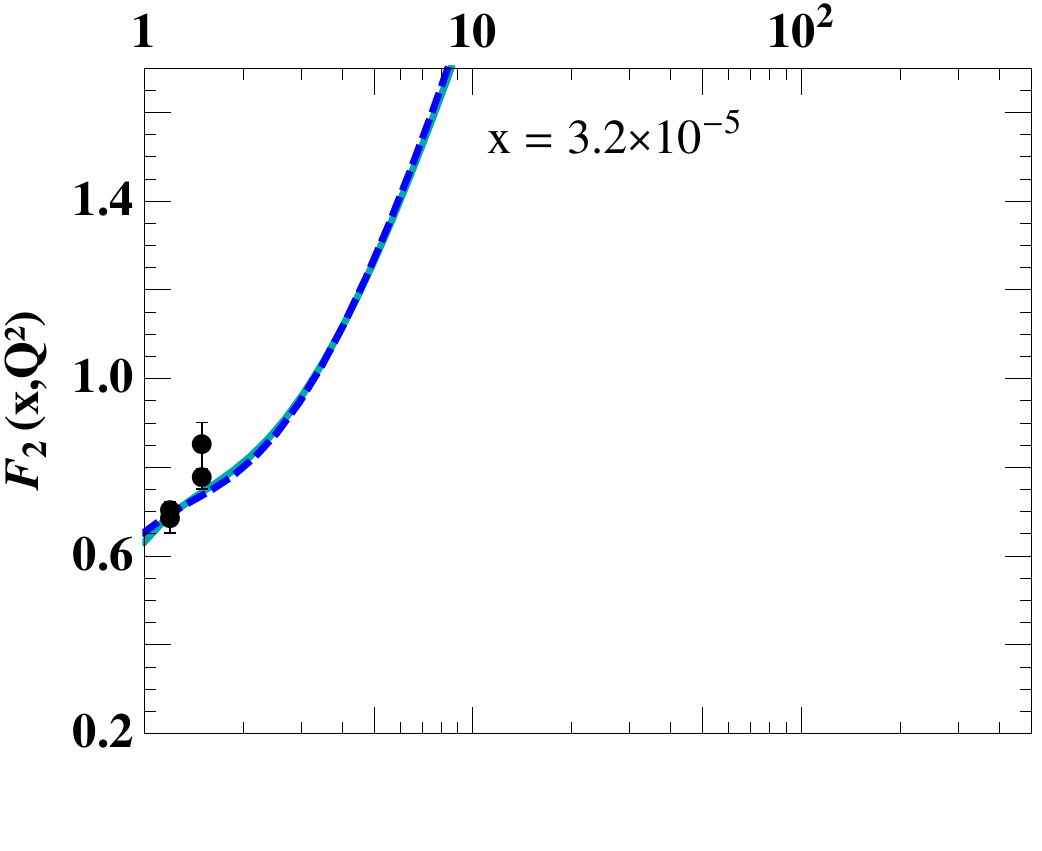} \hspace{-.3cm}
  \includegraphics[scale=.47]{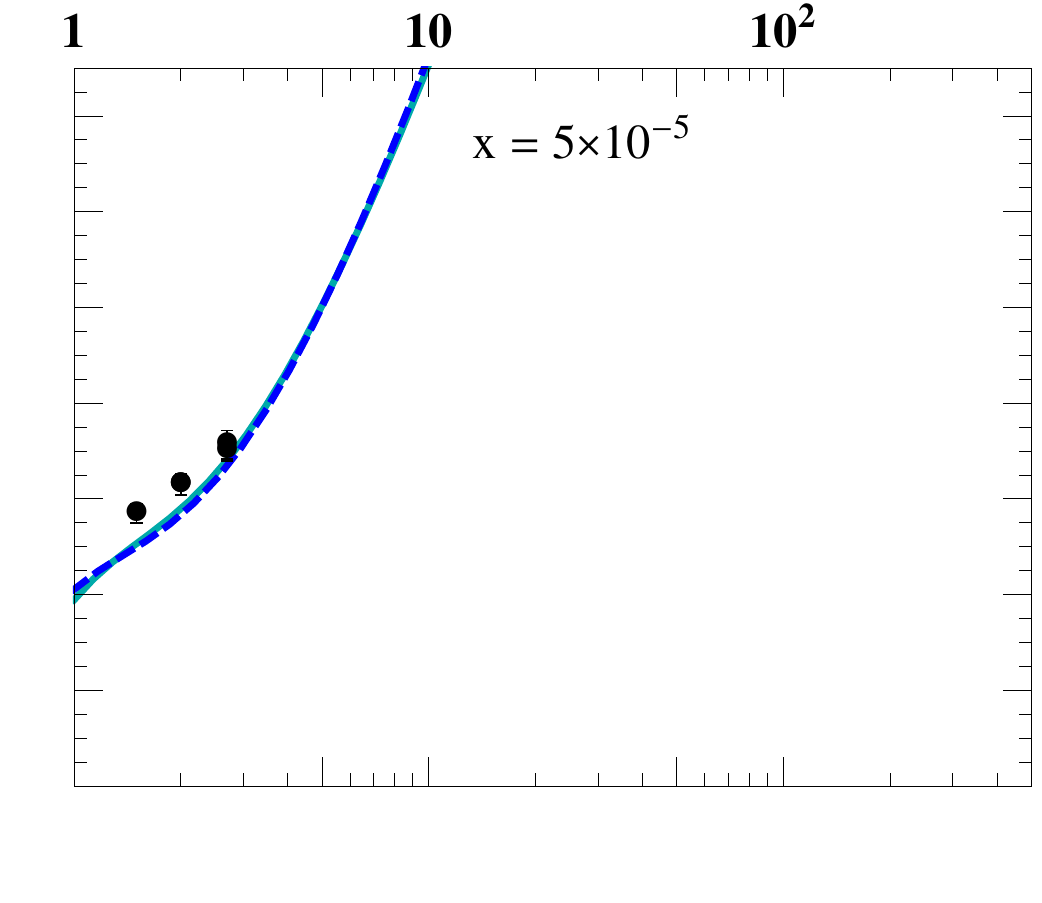}\hspace{-.15cm}
  \includegraphics[scale=.495]{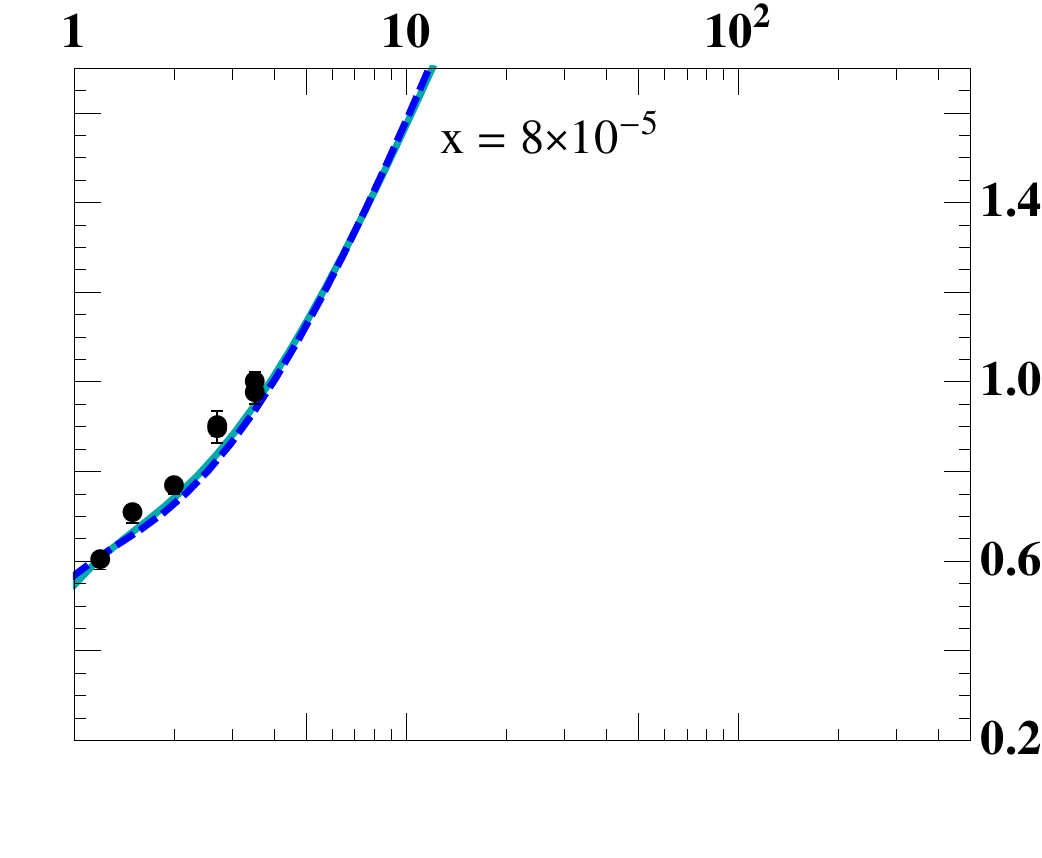} \hspace{-.2cm}
  \hspace{-1.5cm} 
   \vspace{-.3cm} \\
  \hspace{-1.5cm}
   \includegraphics[scale=.5]{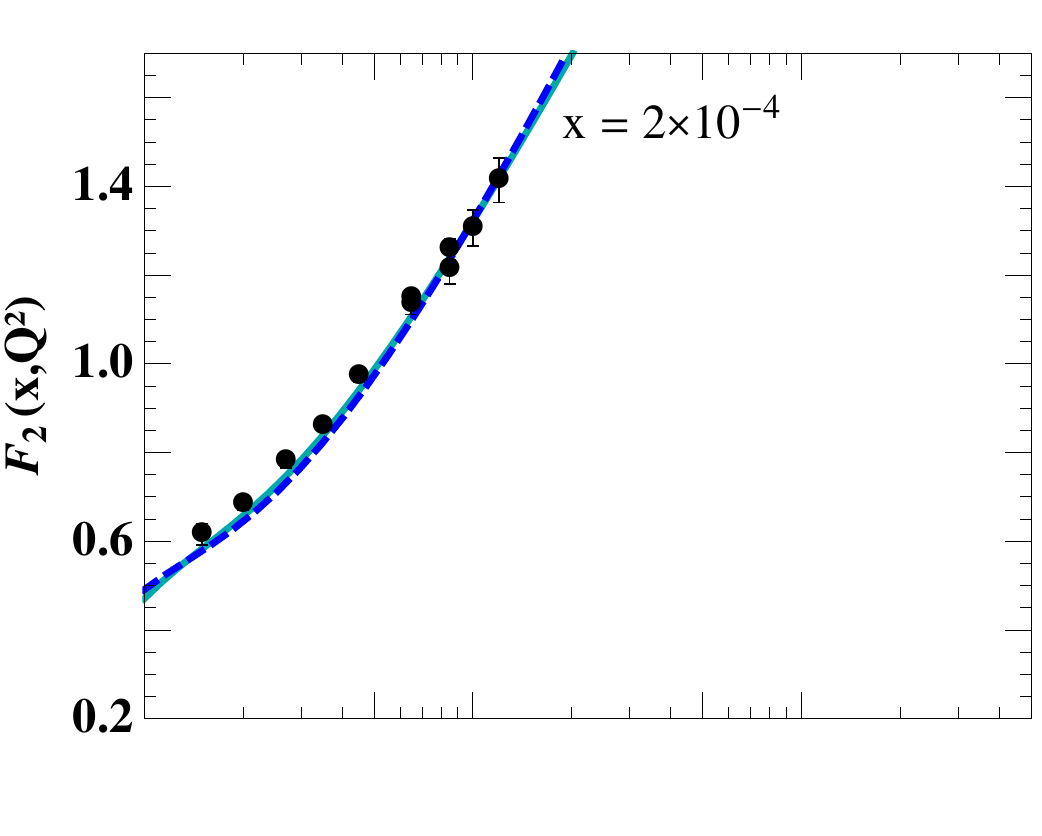} \hspace{-.3cm}
  \includegraphics[scale=.47]{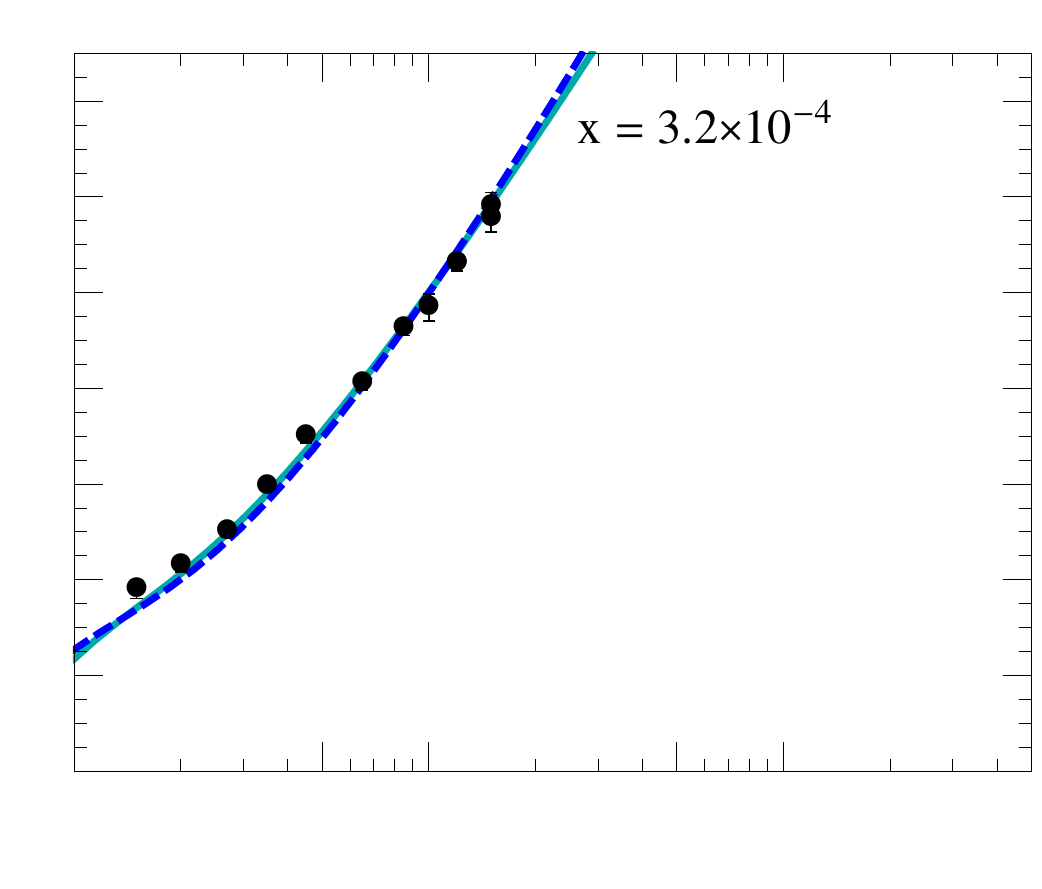}\hspace{-.15cm}
  \includegraphics[scale=.495]{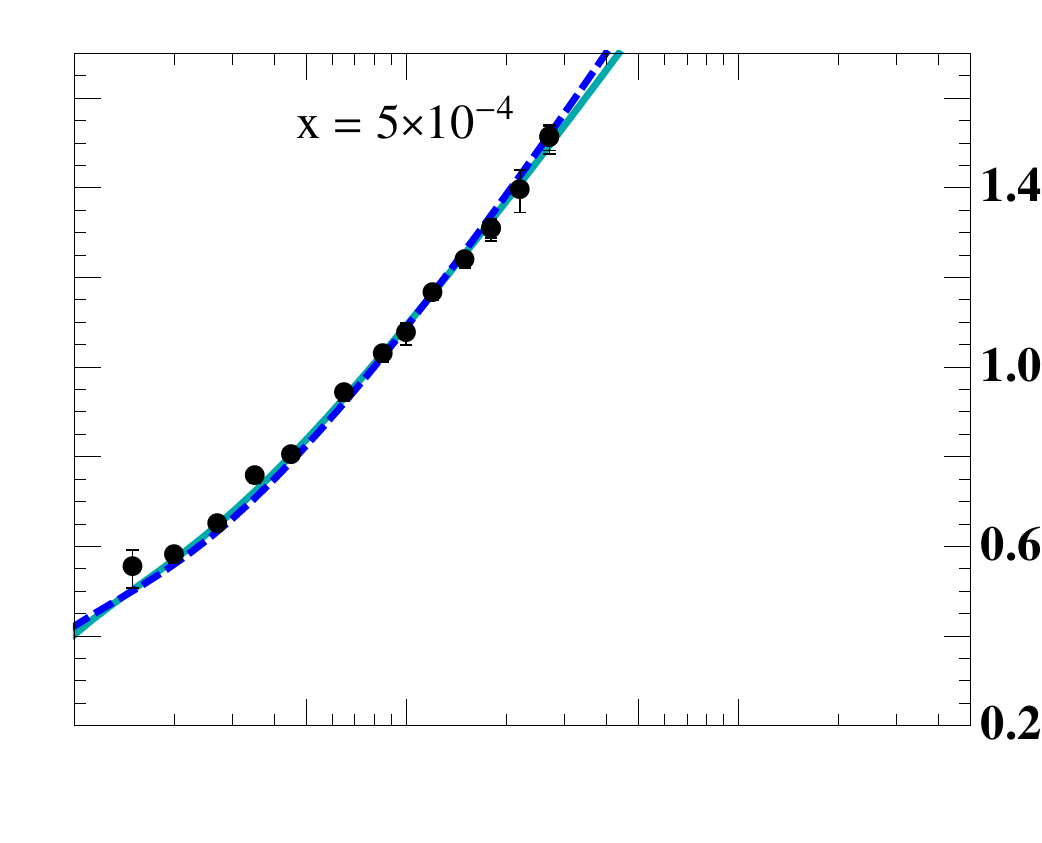} \hspace{-.2cm}
  \hspace{-1.5cm}
   \vspace{-.3cm} \\
  \hspace{-1.5cm}
   \includegraphics[scale=.5]{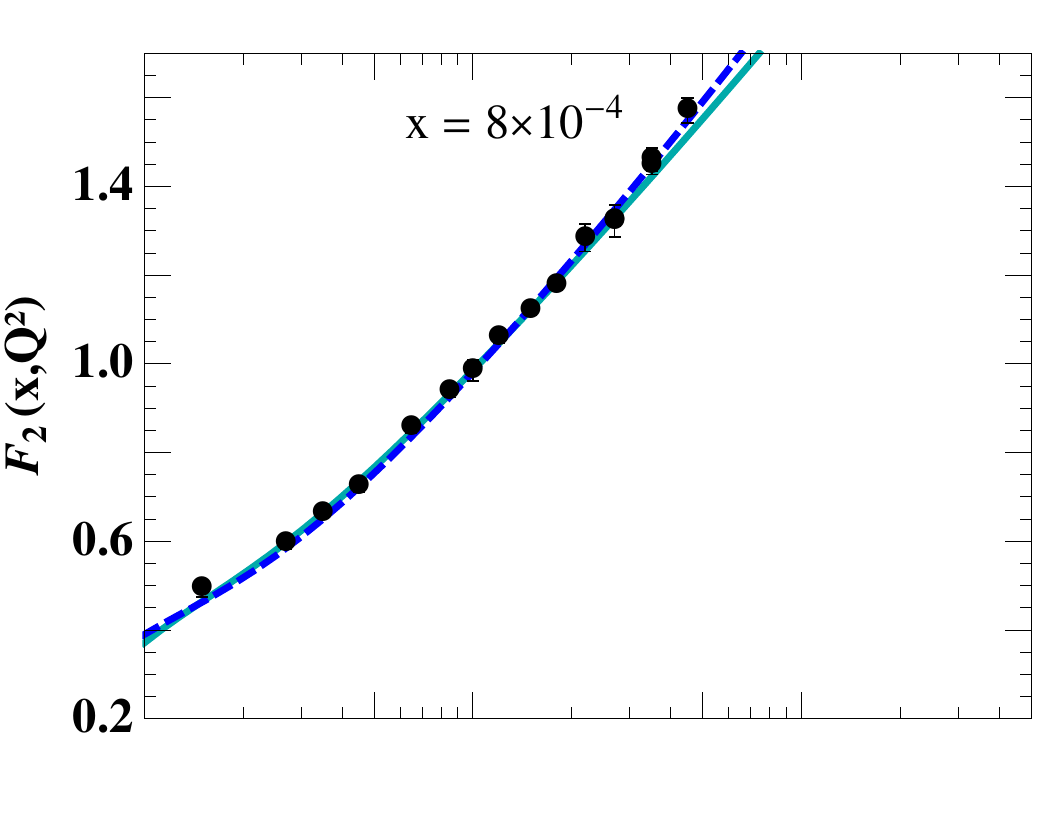} \hspace{-.3cm}
  \includegraphics[scale=.47]{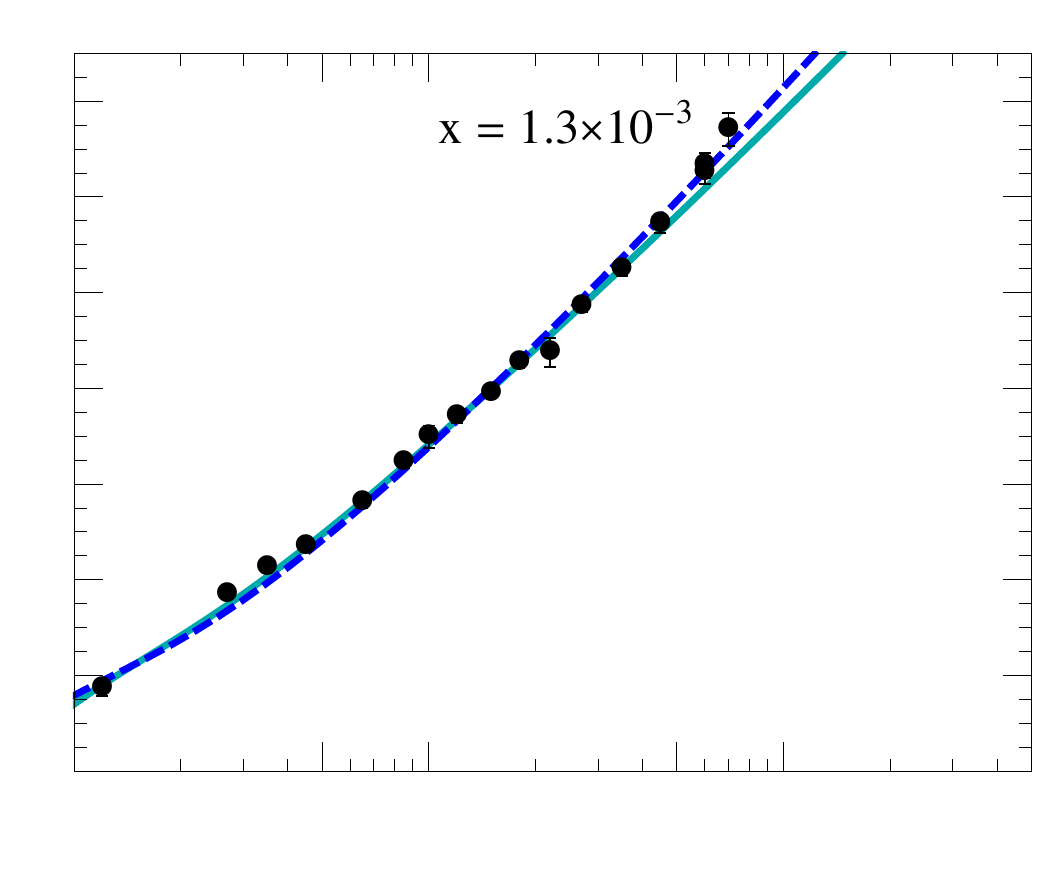}\hspace{-.15cm}
  \includegraphics[scale=.495]{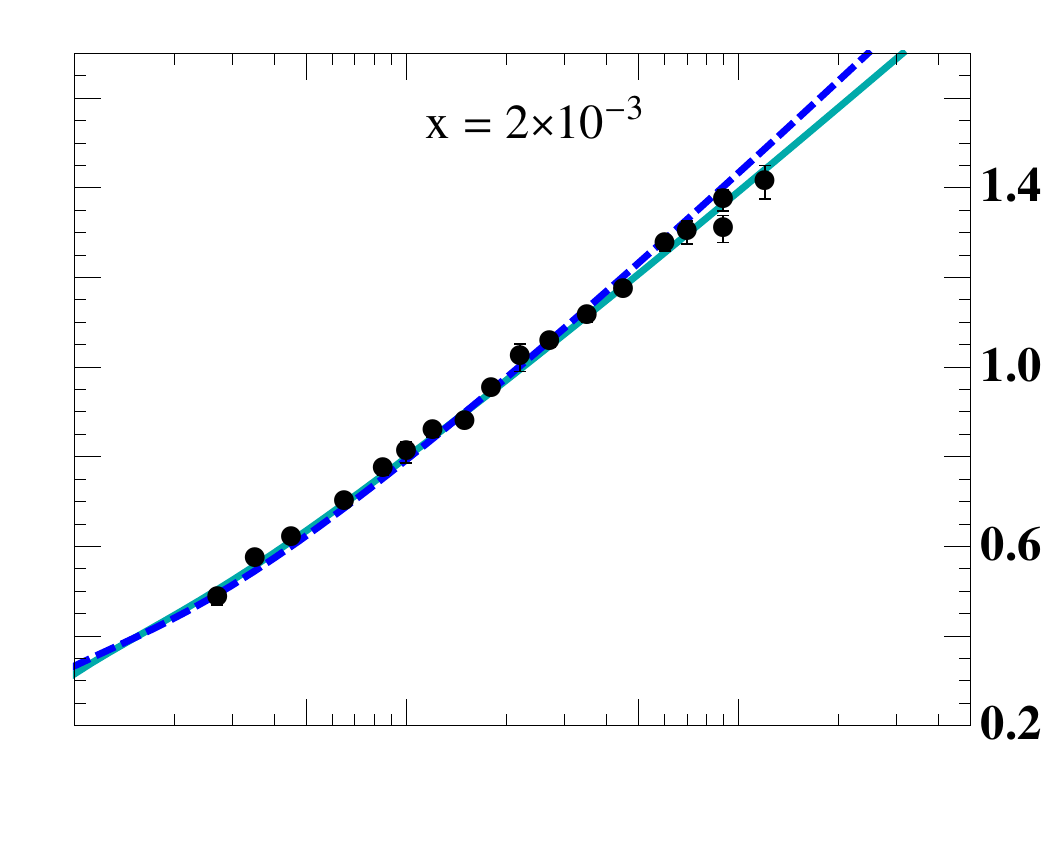}  \hspace{-.2cm}
  \hspace{-1.5cm}
   \vspace{-.3cm} \\
  \hspace{-1.5cm}
   \includegraphics[scale=.5]{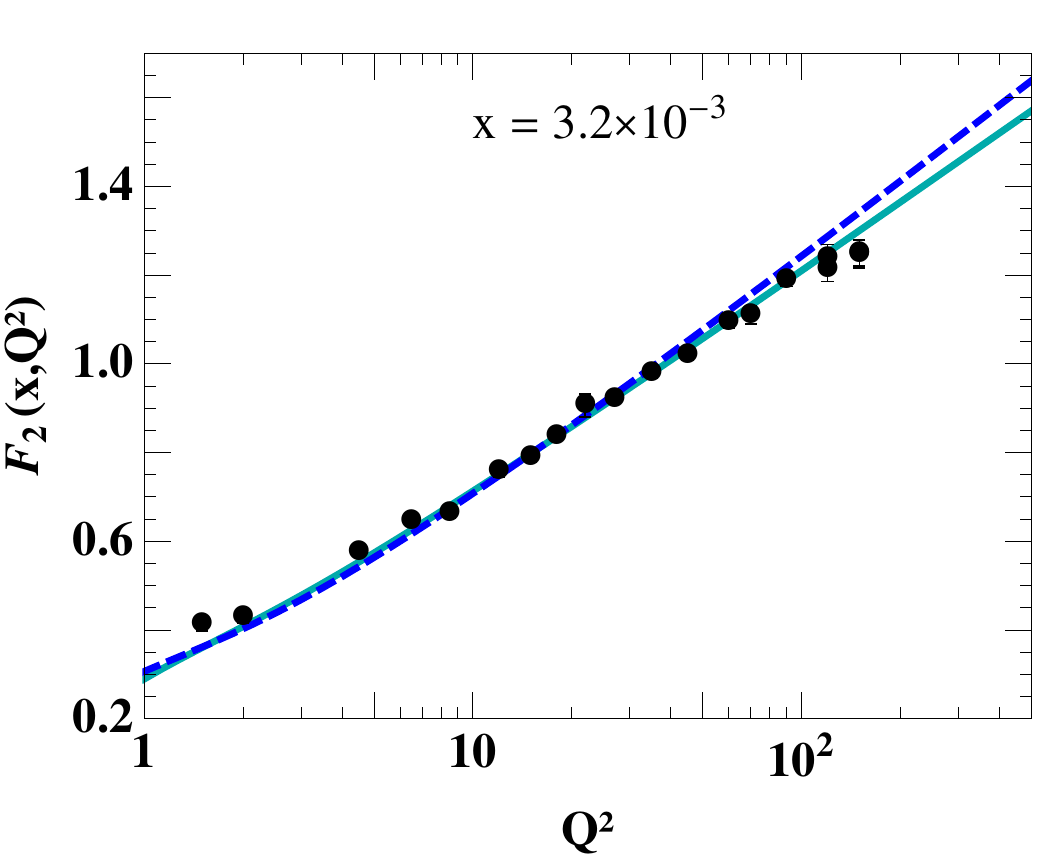} \hspace{-.3cm}
  \includegraphics[scale=.47]{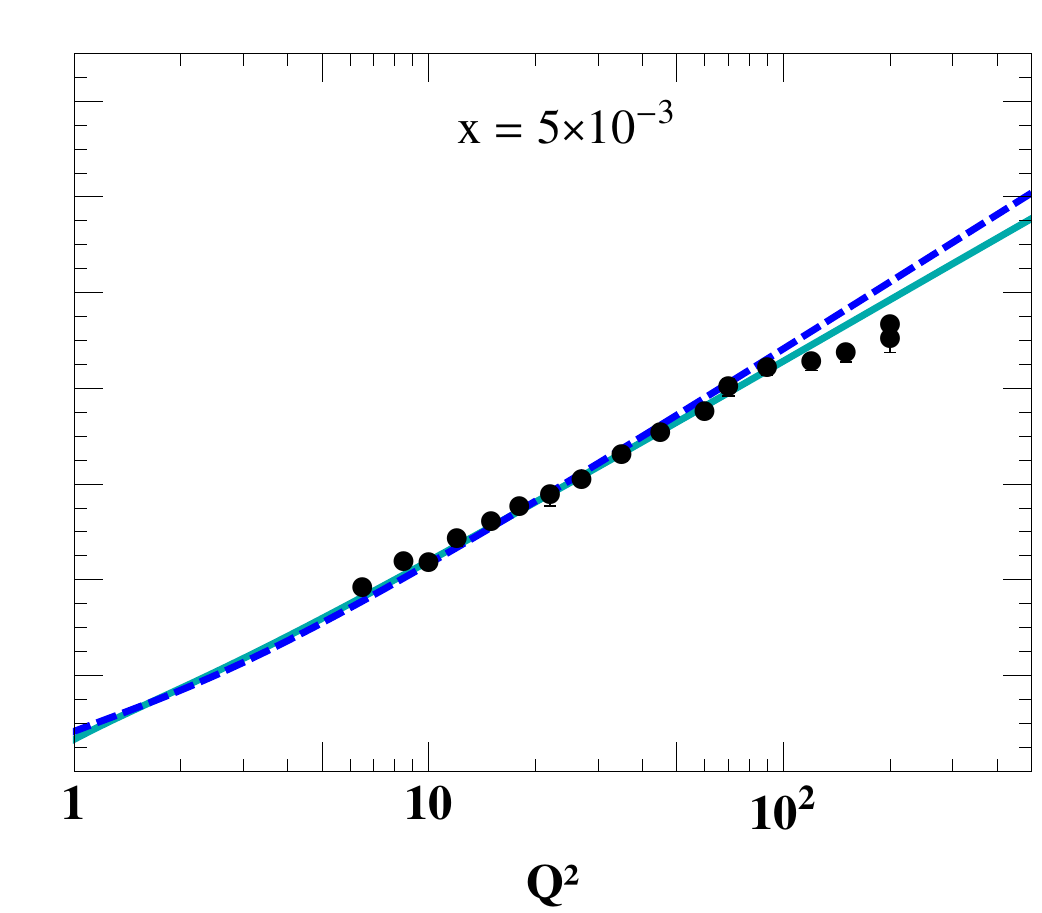}\hspace{-.15cm}
  \includegraphics[scale=.495]{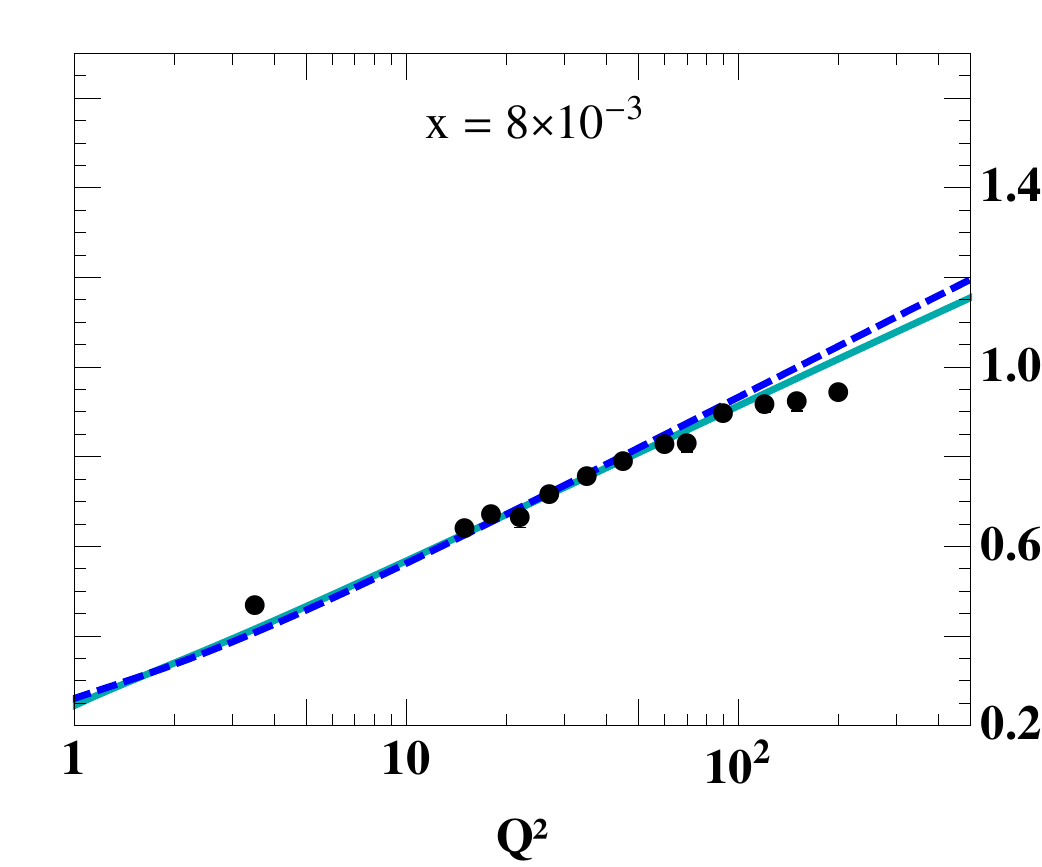} \hspace{-.2cm}
  \hspace{-1.5cm}
  } 
  \caption{Study of the dependence of $F_2 (x,Q^2)$ on $Q^2$ using the LO photon impact factor (solid lines) and the kinematically improved one (dashed lines). $Q^2$ runs from 1.2 to $200\;\text{GeV}^2$.}
  \label{fig:F2}
\end{figure}

\begin{figure}[htbp]
   \centering 
   \hspace{-2cm}
  \includegraphics[scale=.405]{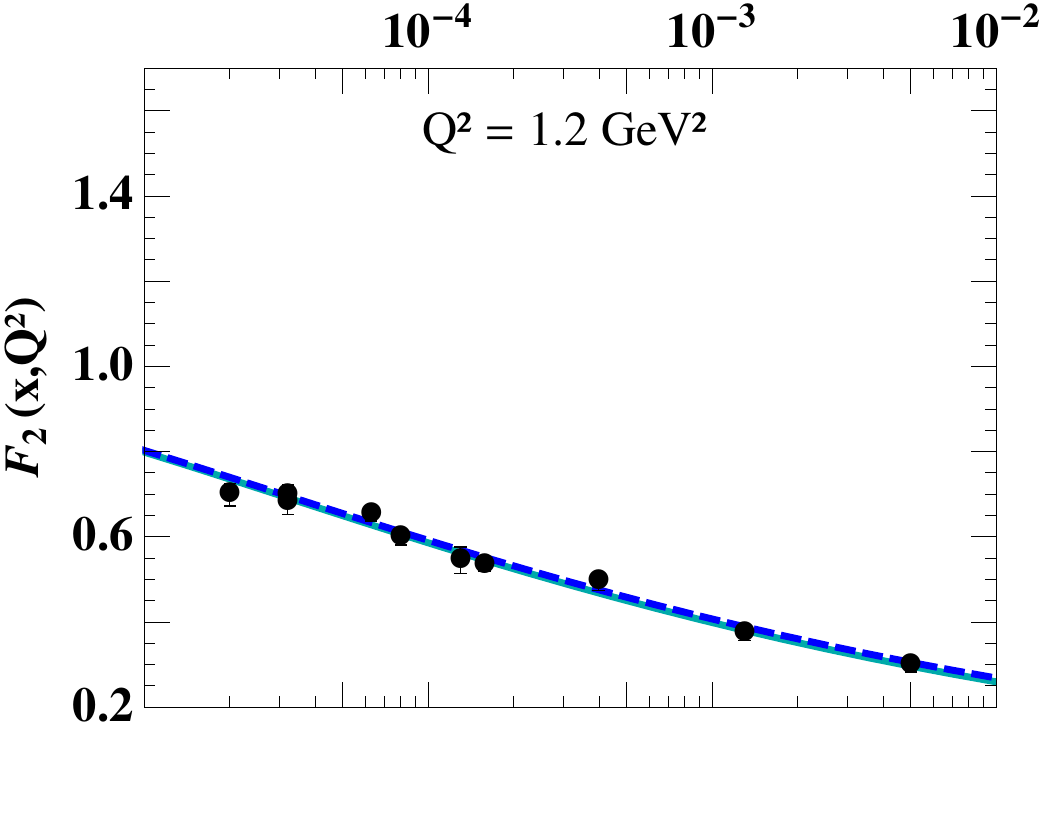} \hspace{-.4cm}
  \includegraphics[scale=.38]{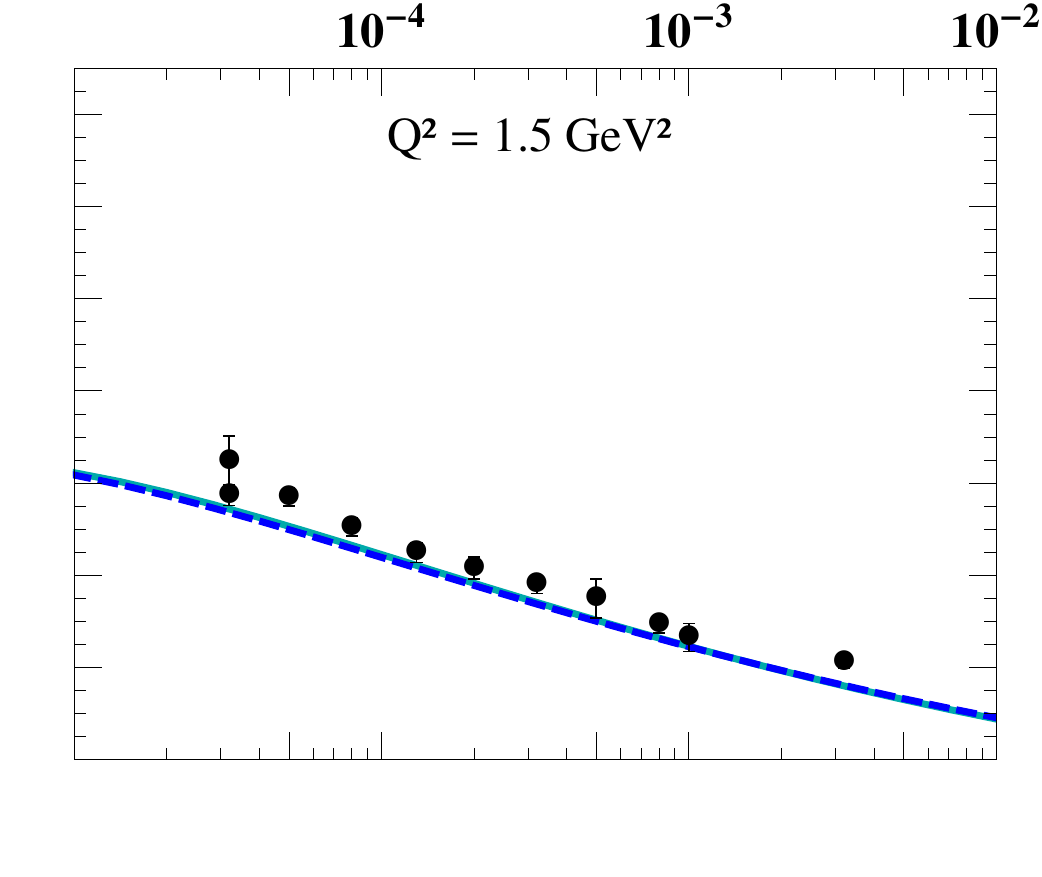}\hspace{-.3cm}
  \includegraphics[scale=.38]{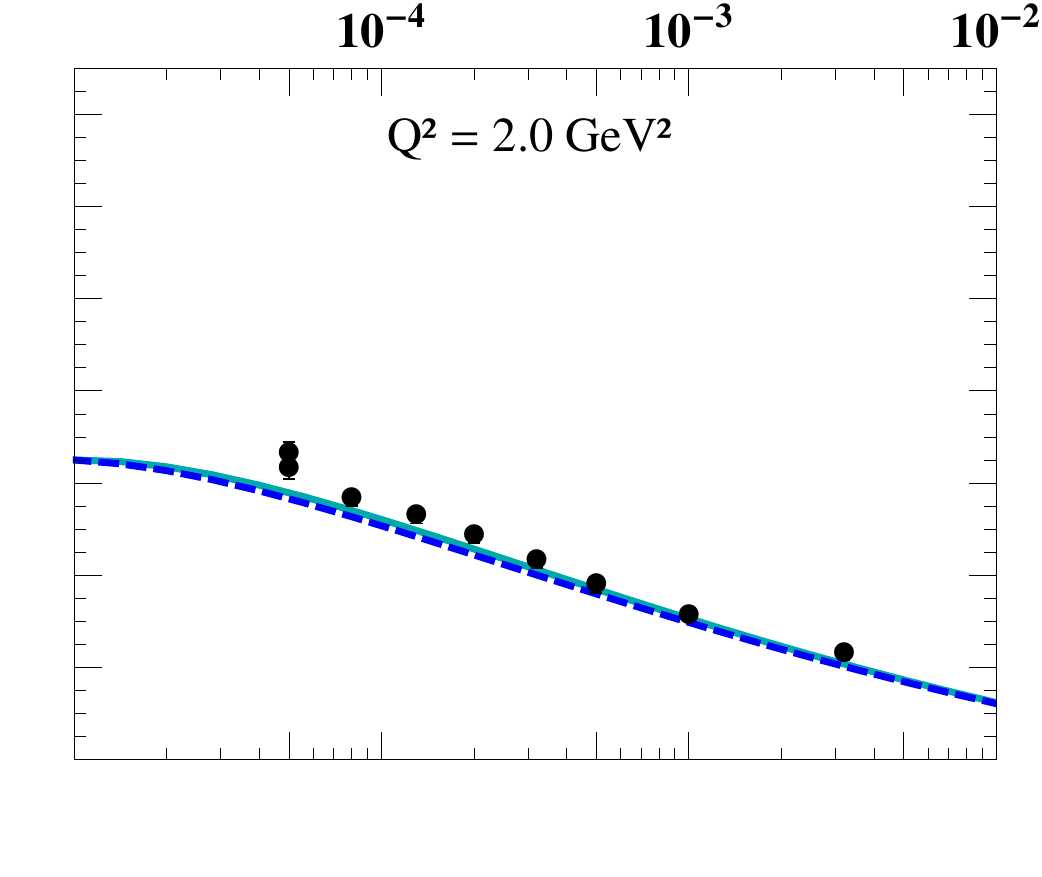} \hspace{-.4cm}
  \includegraphics[scale=.39]{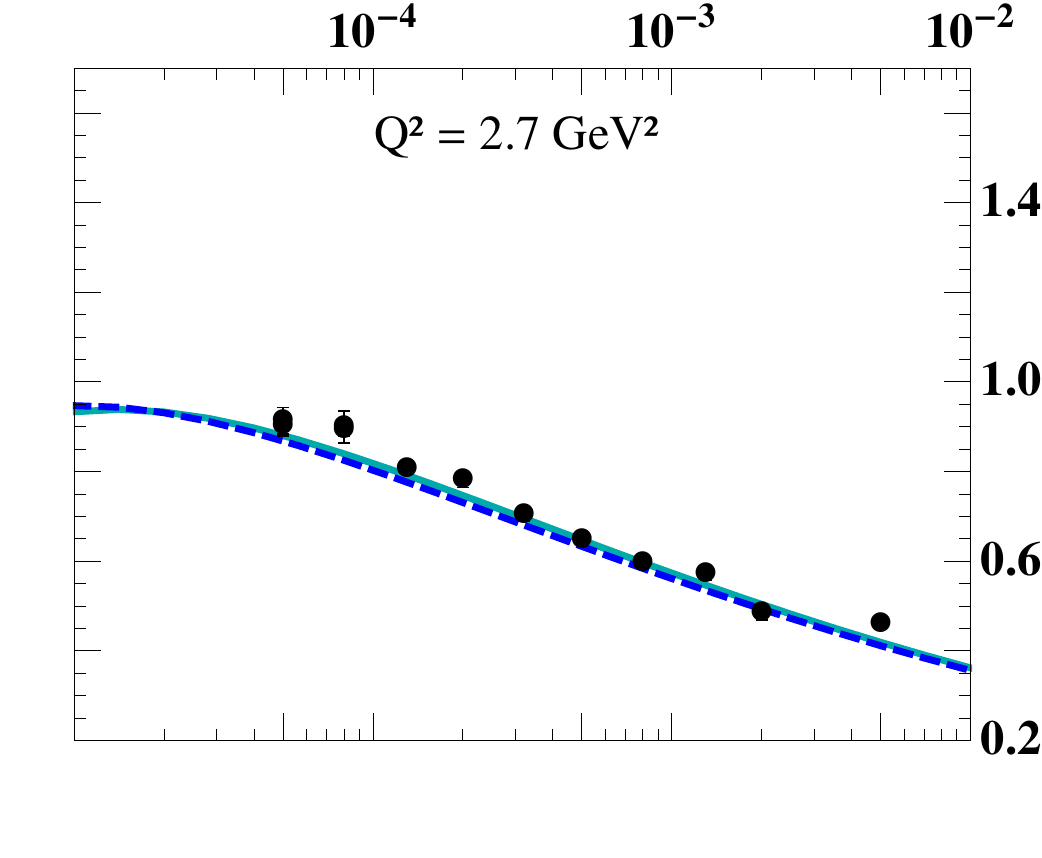} 
   \hspace{-2cm}  \vspace{-.3cm}\\
   \hspace{-2cm}
  \includegraphics[scale=.395]{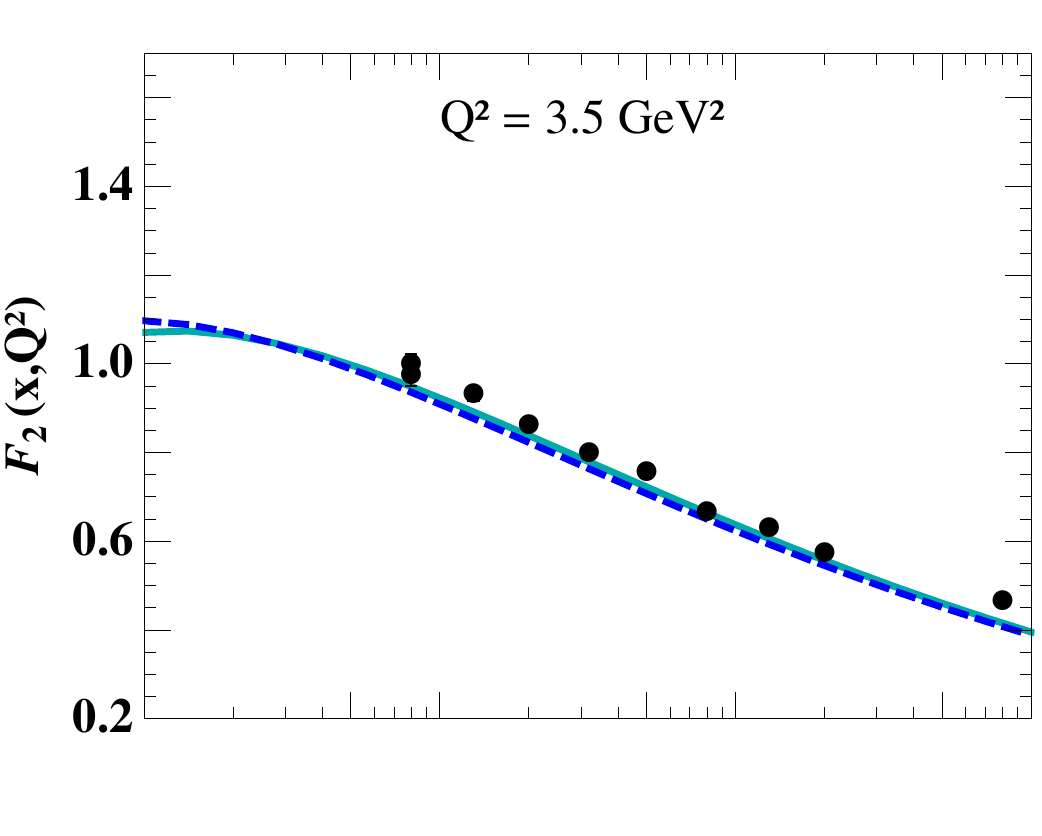} \hspace{-.31cm}
  \includegraphics[scale=.37]{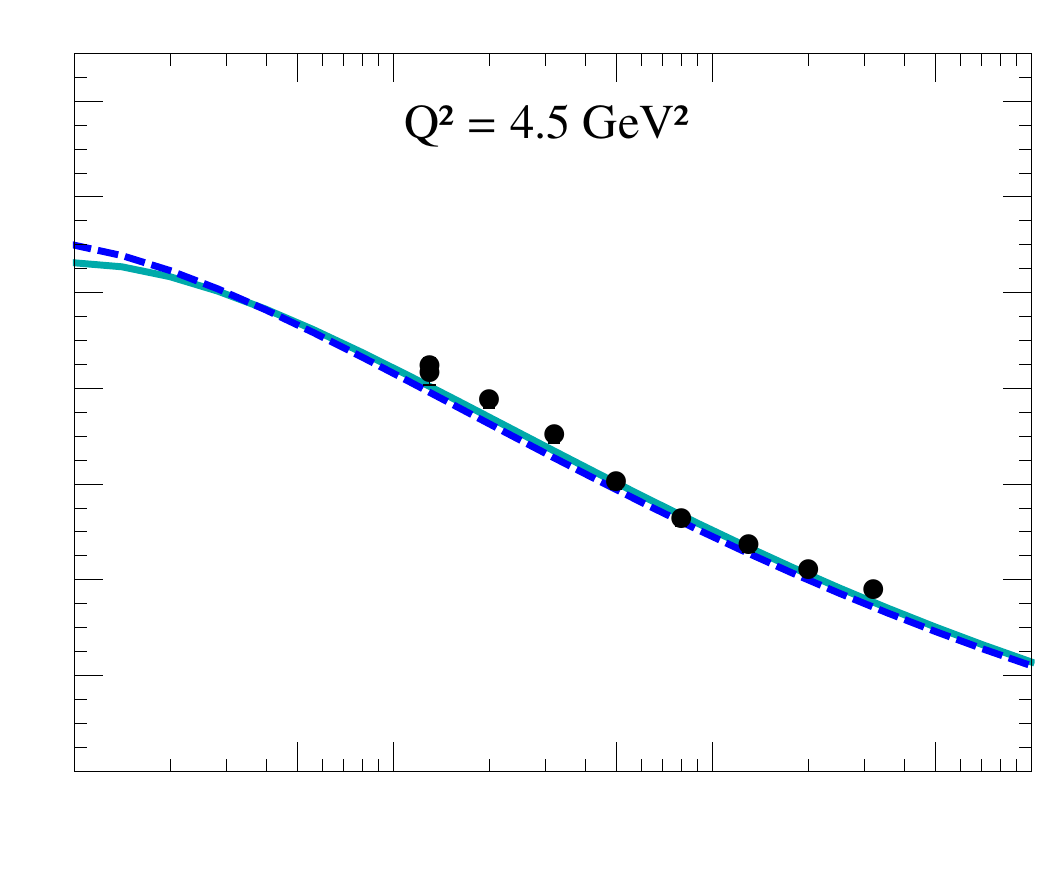}\hspace{-.15cm}
  \includegraphics[scale=.37]{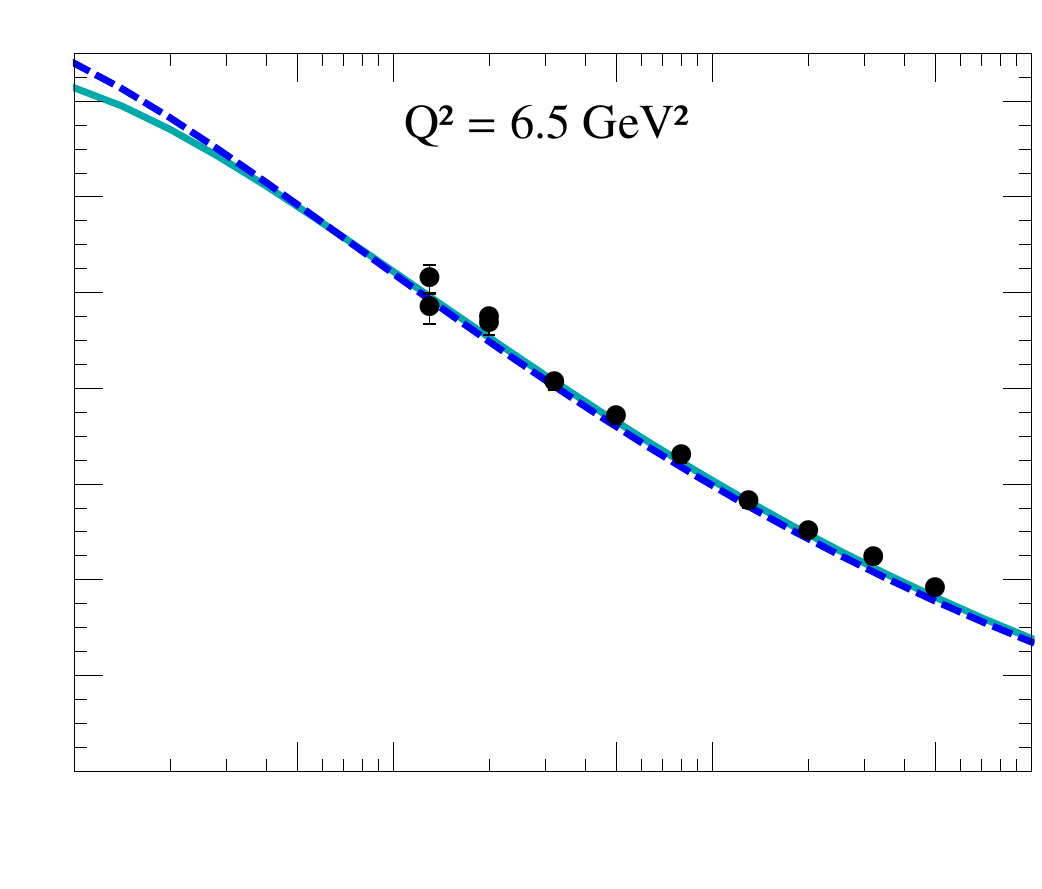} \hspace{-.26cm}
  \includegraphics[scale=.389]{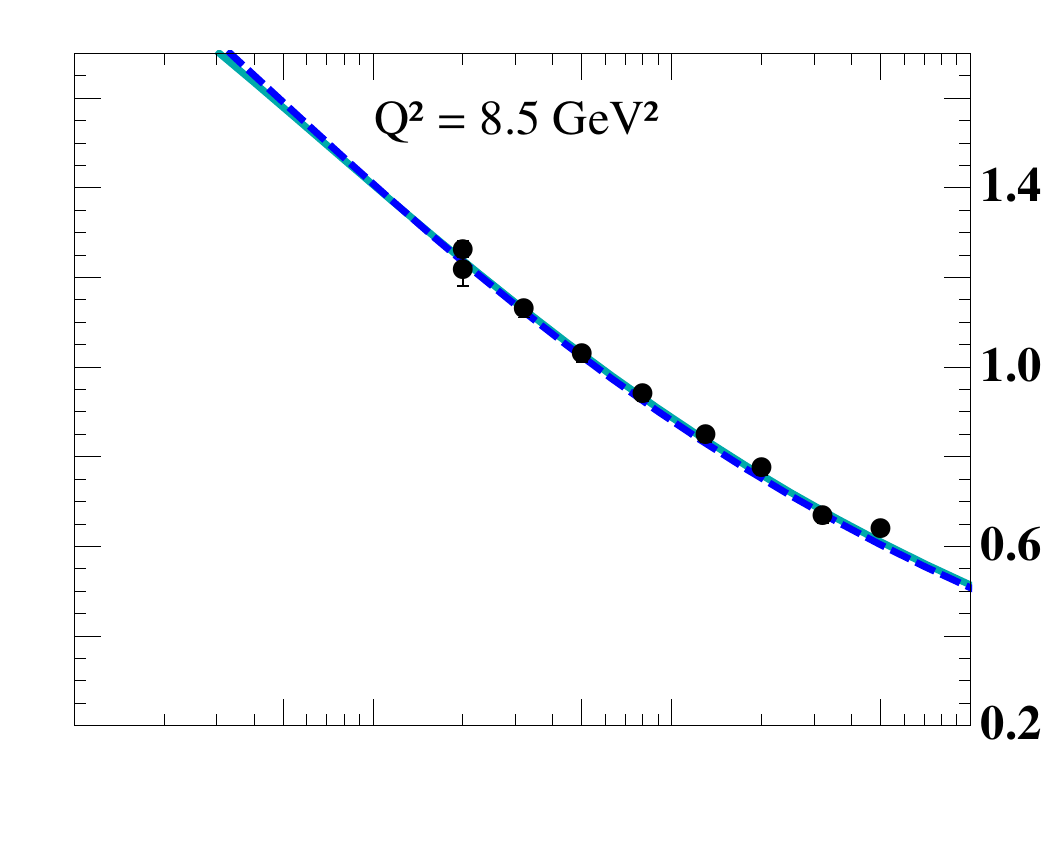} 
  \hspace{-2cm}  \vspace{-.3cm}\\
  \hspace{-2cm}
 \includegraphics[scale=.395]{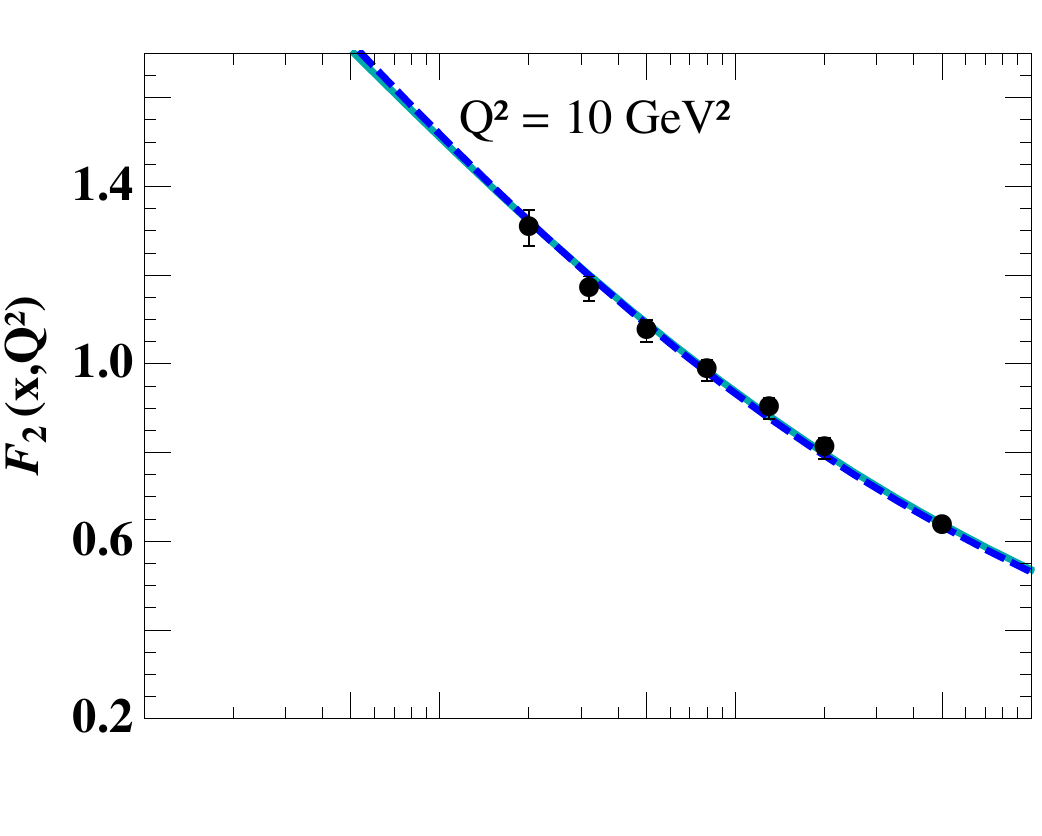} \hspace{-.31cm}
  \includegraphics[scale=.37]{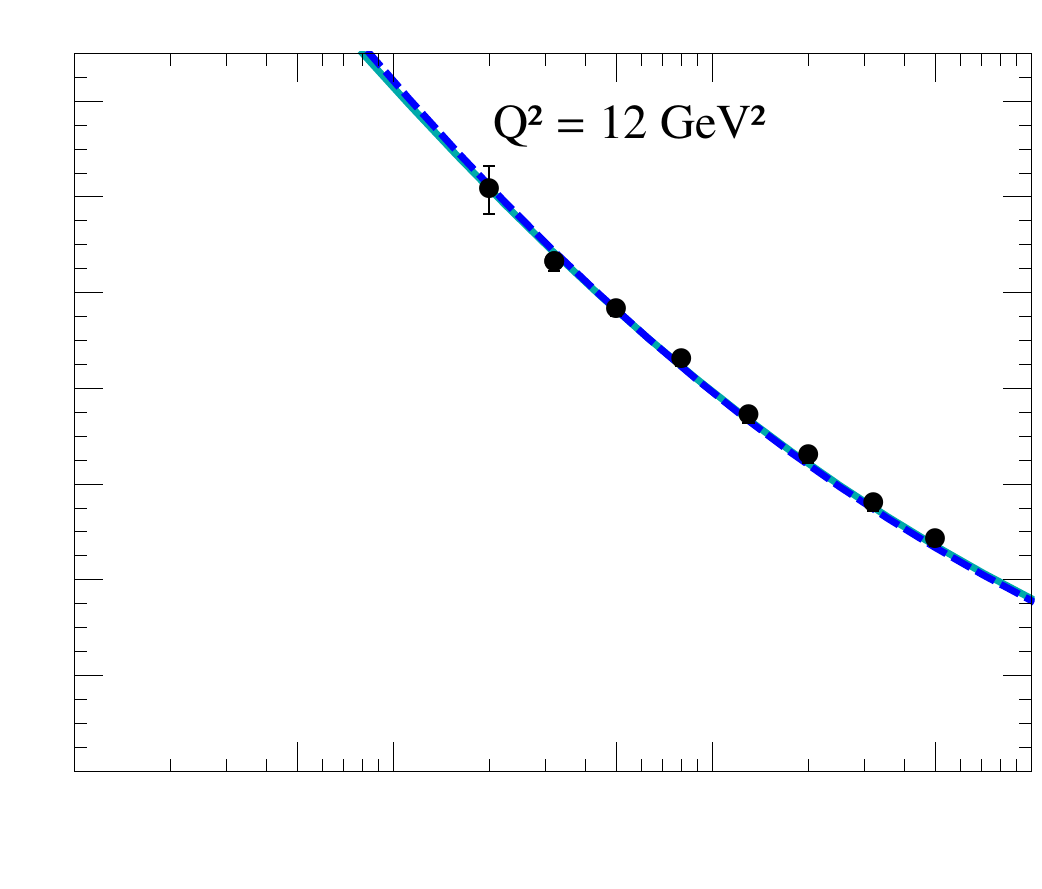}\hspace{-.1cm}
  \includegraphics[scale=.37]{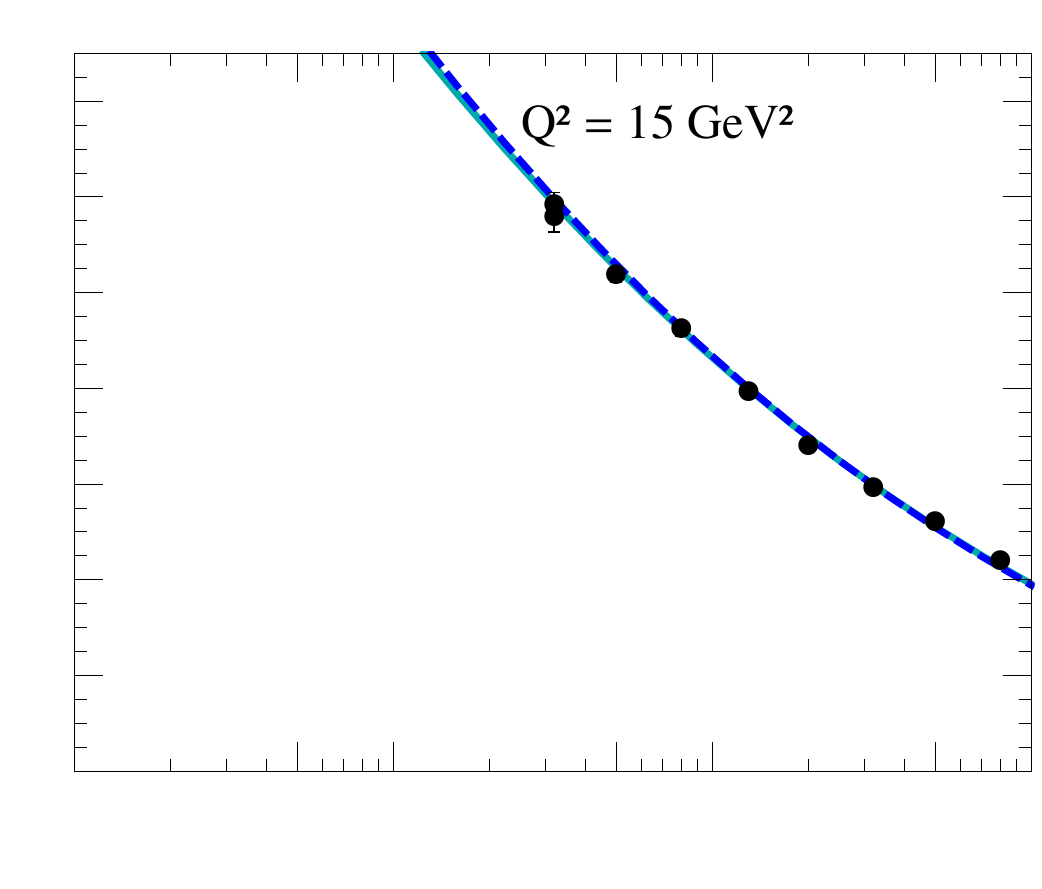} \hspace{-.26cm}
  \includegraphics[scale=.385]{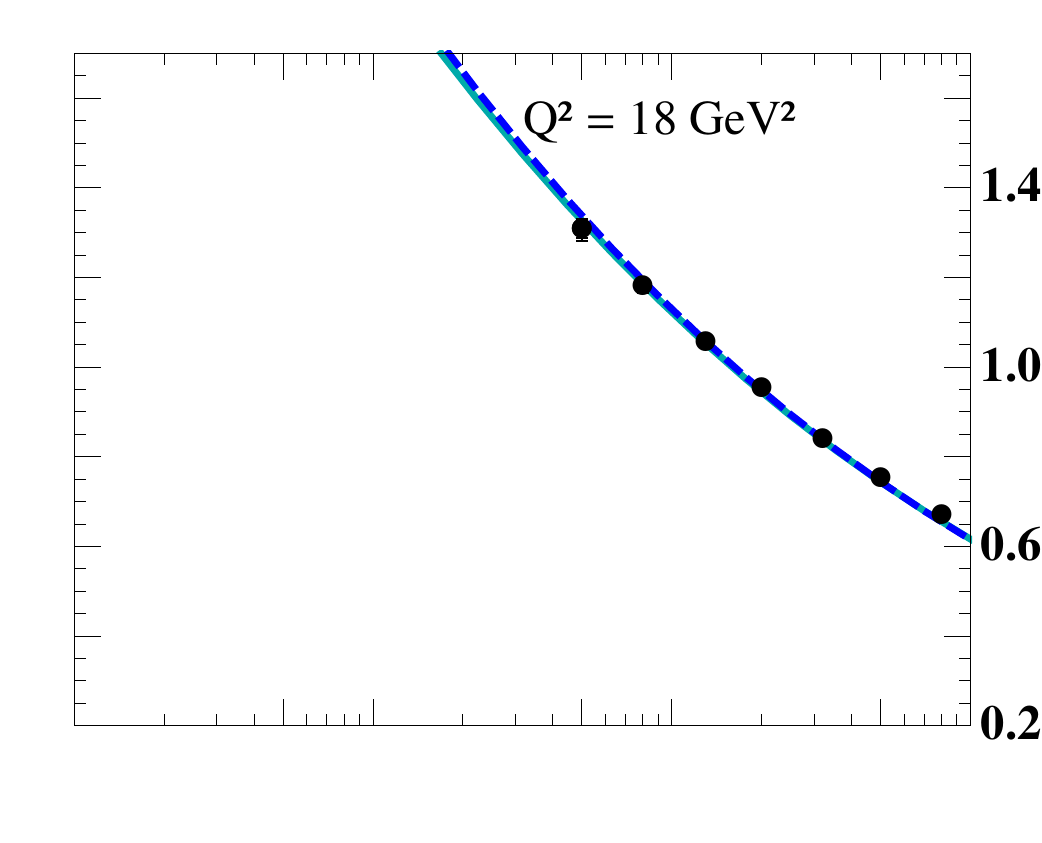} 
  \hspace{-2cm}  \vspace{-.3cm}\\
  \hspace{-2cm}
 \includegraphics[scale=.395]{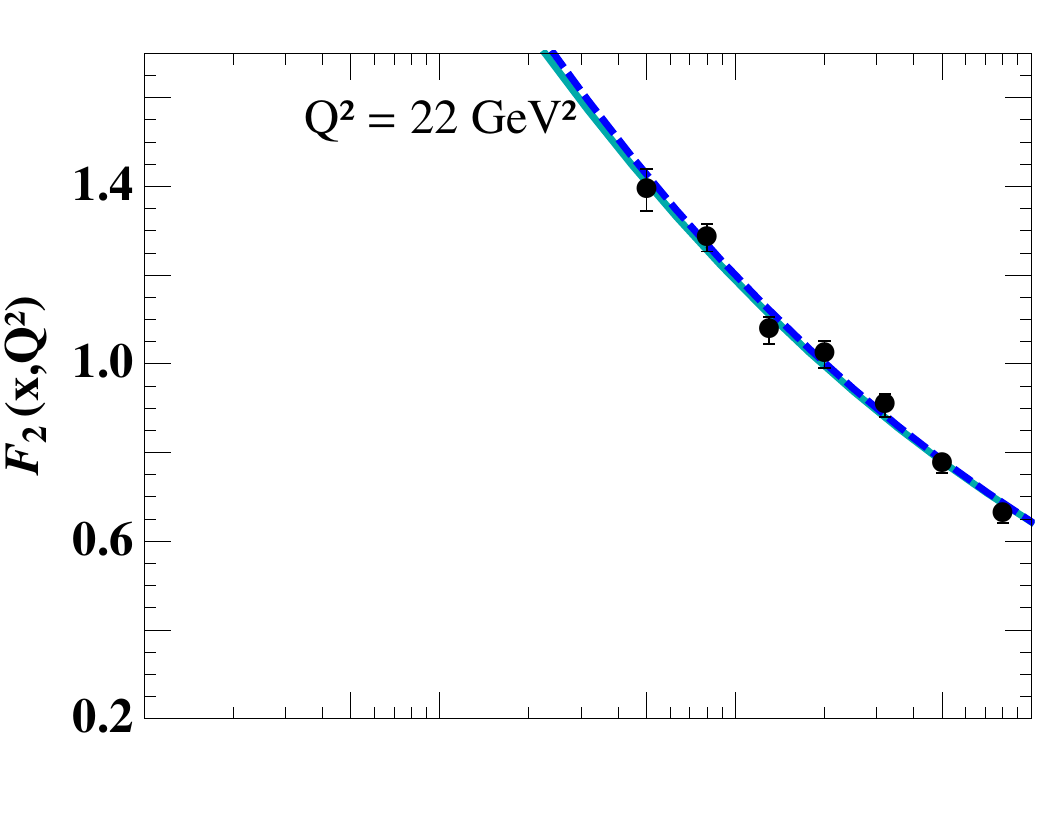} \hspace{-.32cm}
  \includegraphics[scale=.37]{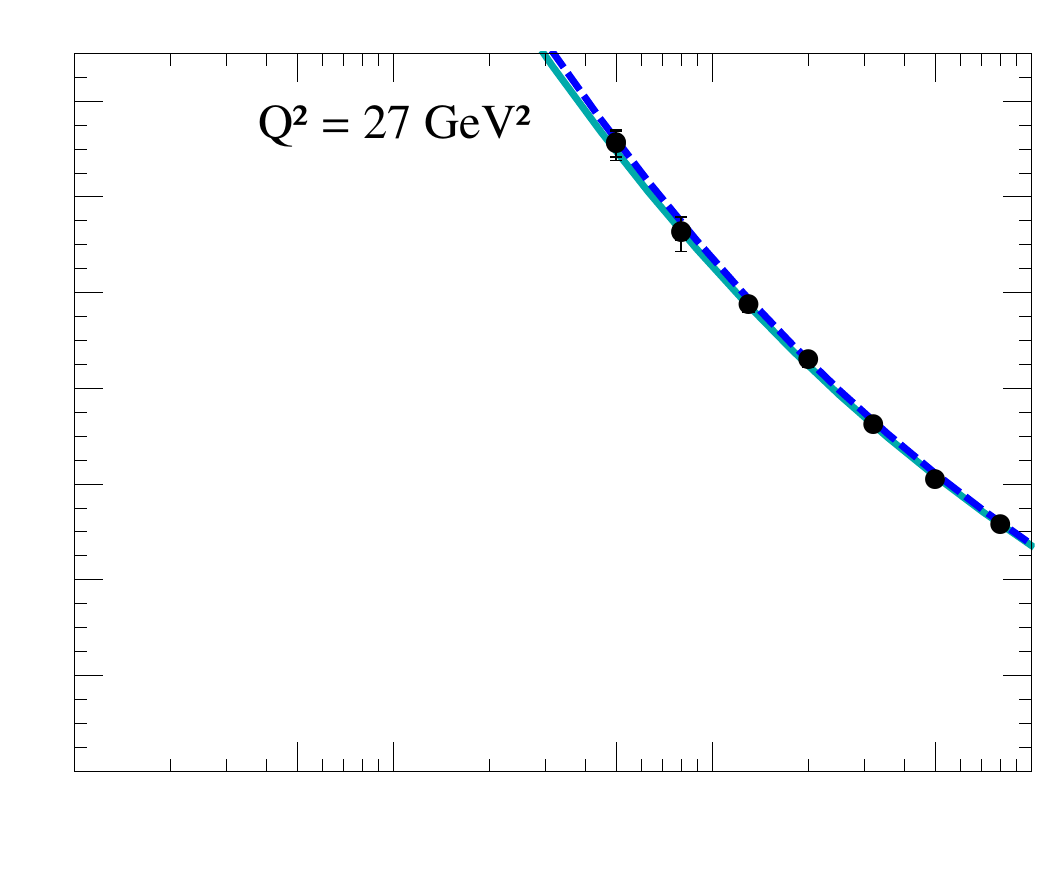}\hspace{-.15cm}
  \includegraphics[scale=.37]{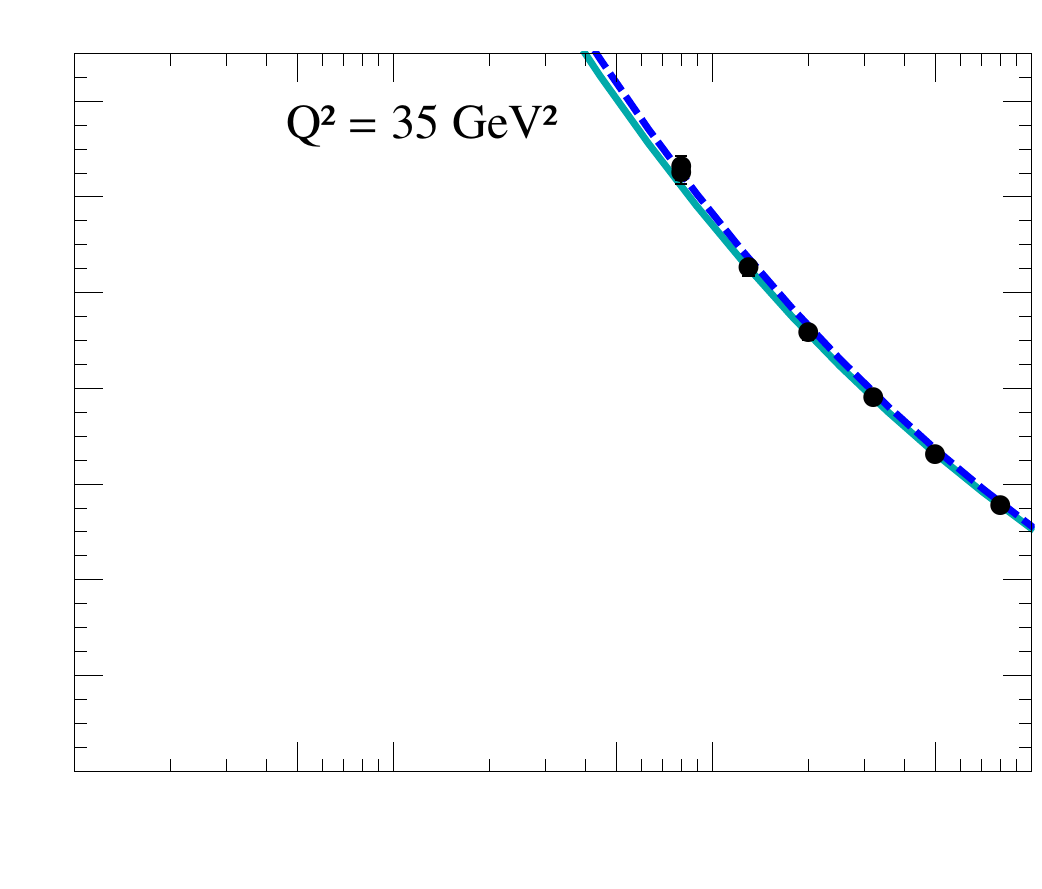} \hspace{-.26cm}
  \includegraphics[scale=.389]{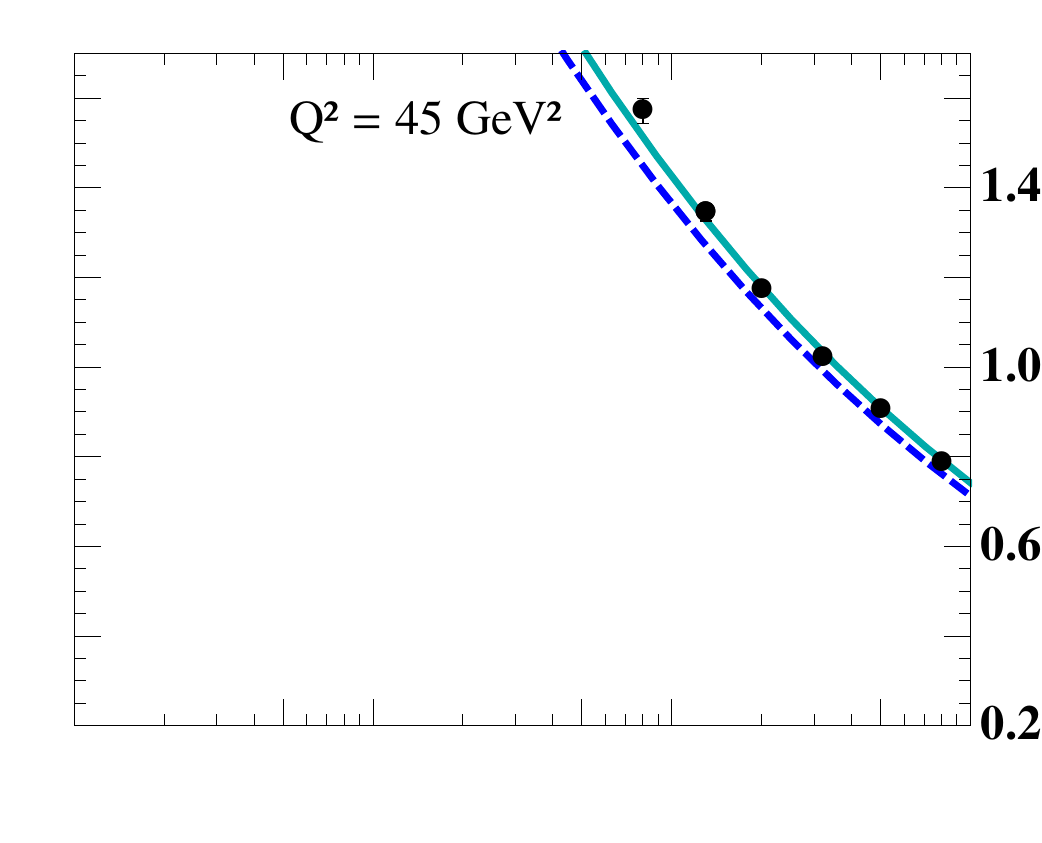} 
  \hspace{-2cm}
   \vspace{-.3cm}  \\
  \hspace{-2cm}
   \includegraphics[scale=.4]{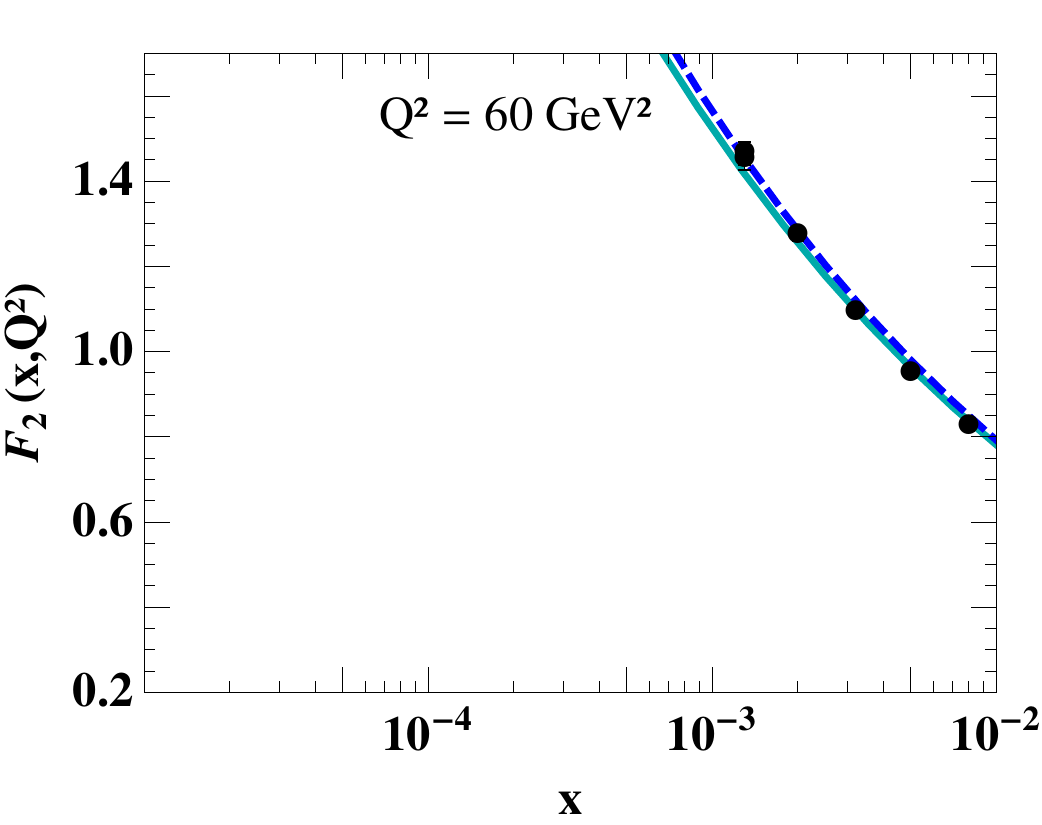} \hspace{-.42cm}
  \includegraphics[scale=.38]{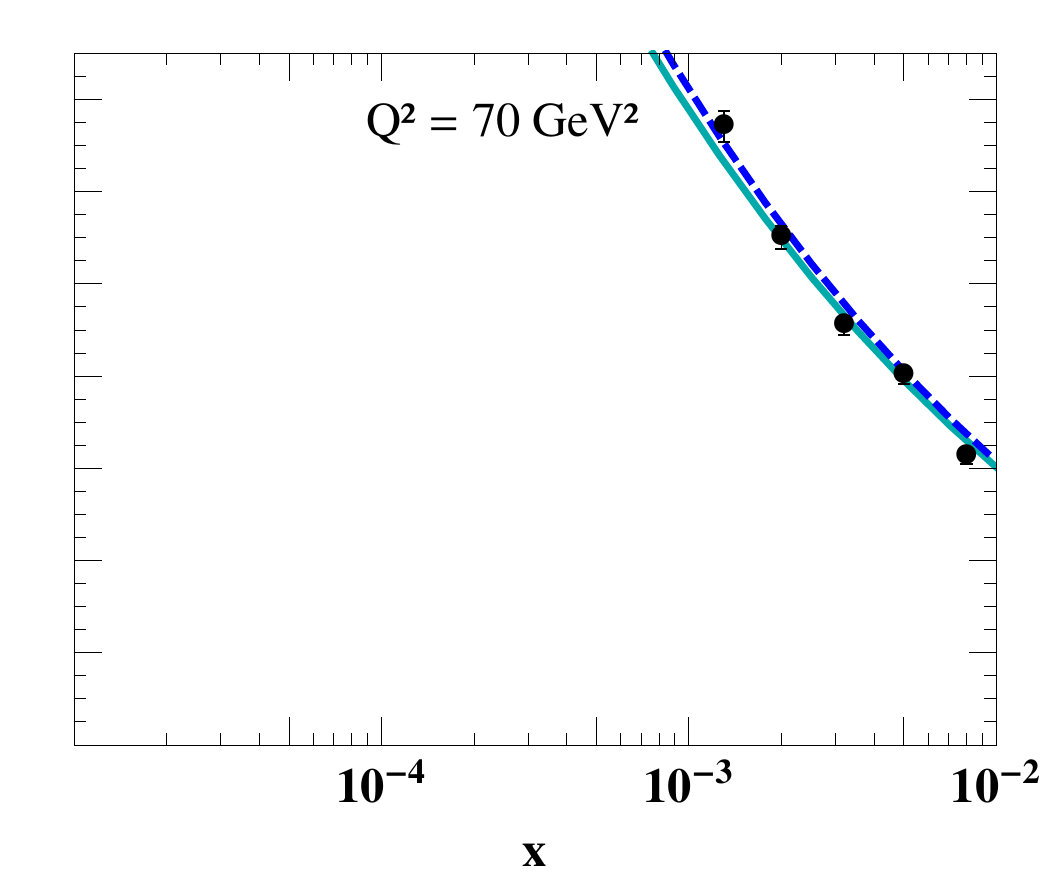}\hspace{-.25cm}
  \includegraphics[scale=.38]{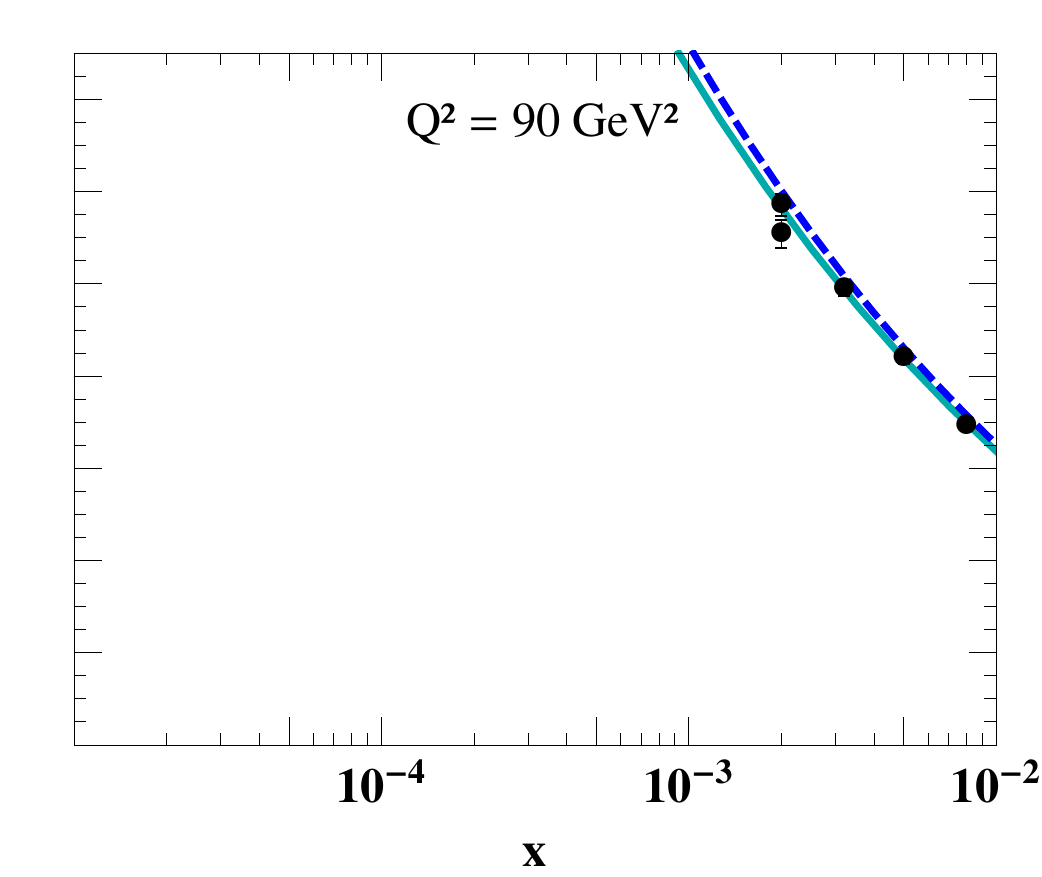} \hspace{-.36cm}
  \includegraphics[scale=.39]{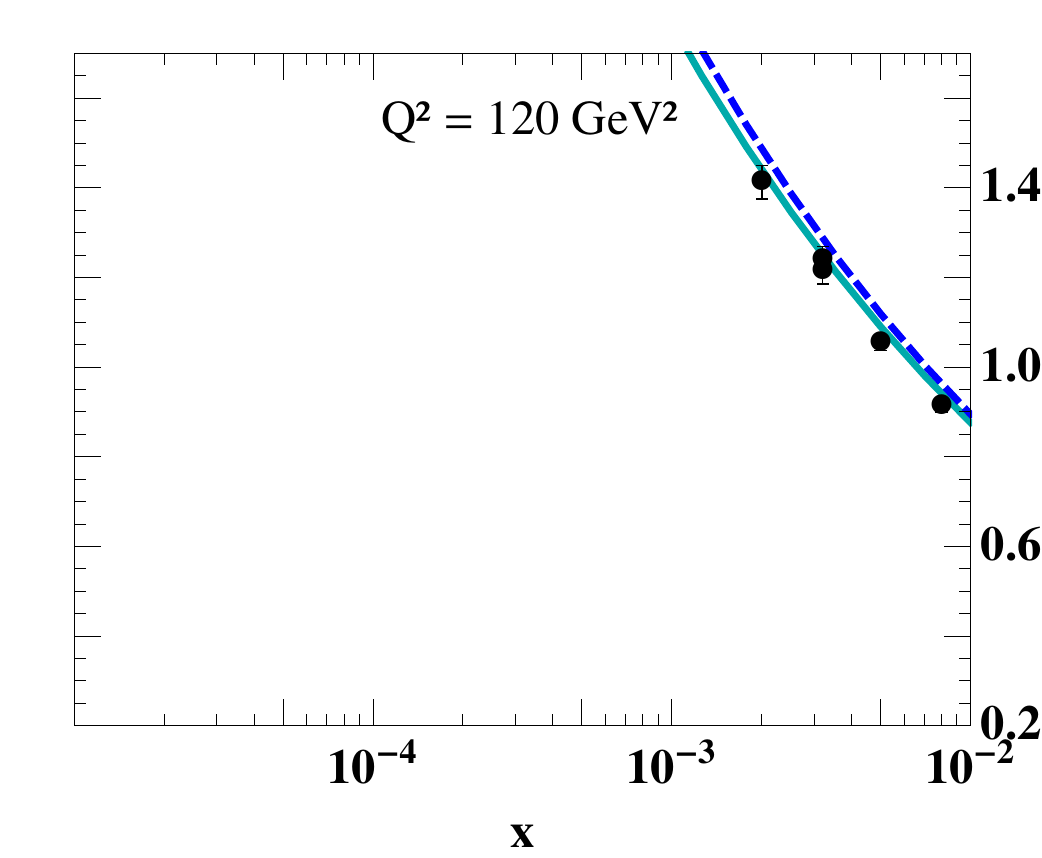} 
  \hspace{-2cm}
  %\\
  % \includegraphics[scale=.35]{F2150BLM.pdf} 
  % \includegraphics[scale=.35]{F2200BLM.pdf} 
  \caption{Study of the dependence of $F_2 (x,Q^2)$ on $x$ using the LO photon impact factor (solid lines) and the kinematically improved one (dashed lines). $Q^2$ runs from 1.2 to $120 \;\text{GeV}^2$.}
  \label{fig:f2x}
\end{figure}

Encouraged by these positive results we now turn to investigate more differential distributions. We select data with fixed 
values of $x$ and compare the $Q^2$ dependence of our theoretical predictions with them, now fixing the normalization 
for the LO impact factor to ${\cal C} = 1.50$ and $2.39$ for the kinematically 
improved.  Our results are presented in Fig.~\ref{fig:F2}.  
The equivalent comparison to data, this time fixing $Q^2$ and looking into the evolution in the $x$ variable, is shown 
in Fig.~\ref{fig:f2x}. 
We observe that our predictions give 
a very accurate description of the data for both types of impact factors.

 Let us remark that the values for the parameters 
in this fit are in syntony with the theoretical expectations for the proton impact factor since $Q_0$ is very similar to the confinement scale and the value of $\delta$ sets the maximal contribution from the impact factor also in that region. This 
is reasonable given that the proton has a large transverse size.

\subsection{$F_L$}

\begin{figure}[htbp]
  \centering \hspace{-2cm}
  \includegraphics[scale=.52]{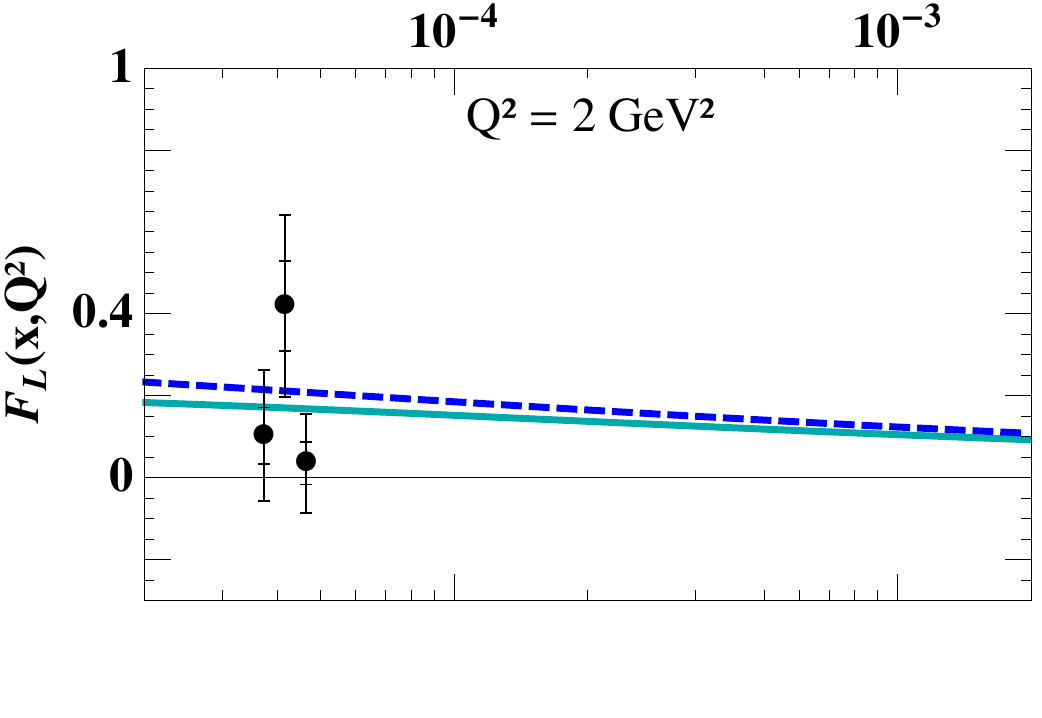} \hspace{-.33cm}
  \includegraphics[scale=.49]{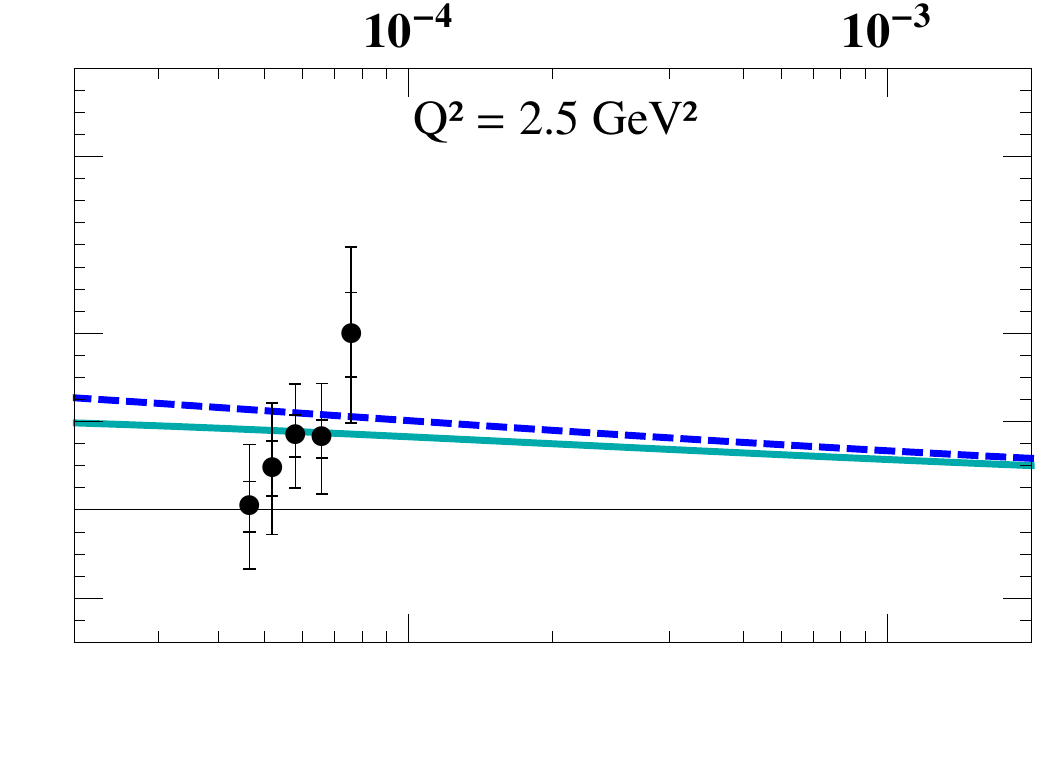}\hspace{-.17cm}
  \includegraphics[scale=.51]{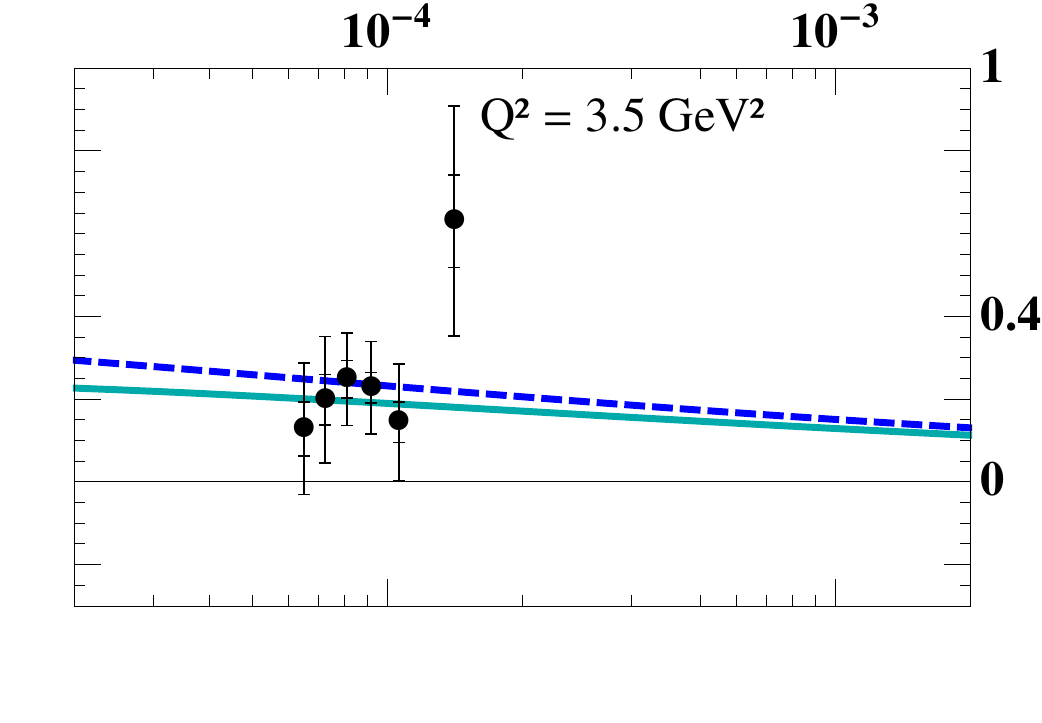}   \hspace{-2cm} \vspace{-.3cm}\\
  \hspace{-2cm}
  %\vspace{-.2cm}
  \includegraphics[scale=.52]{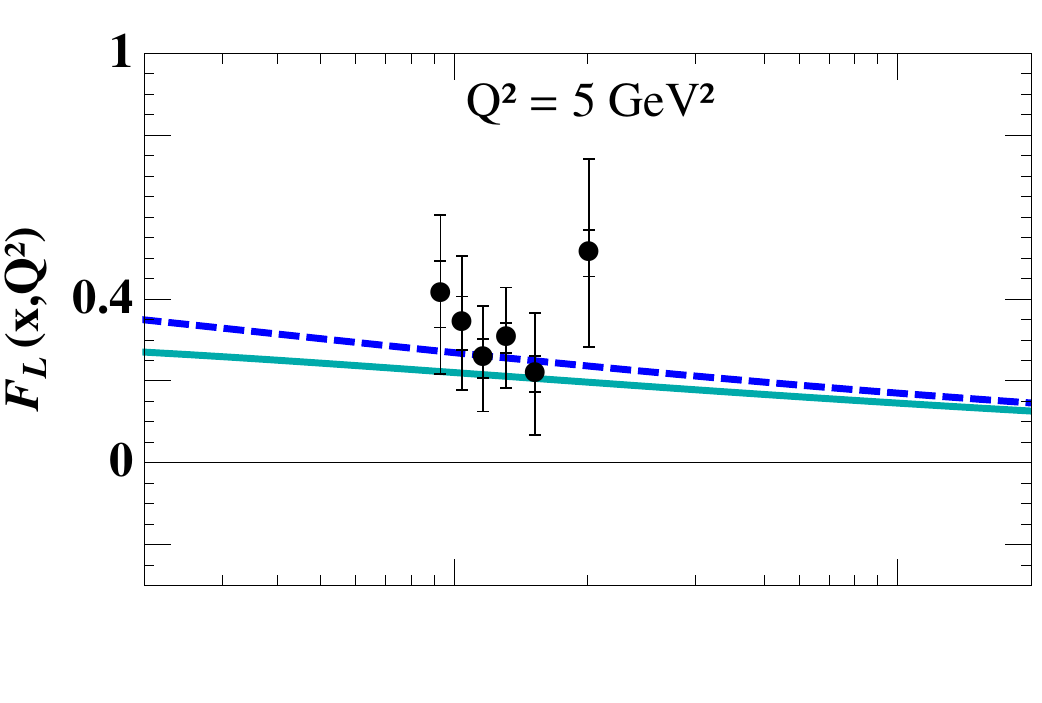} \hspace{-.33cm}
  \includegraphics[scale=.49]{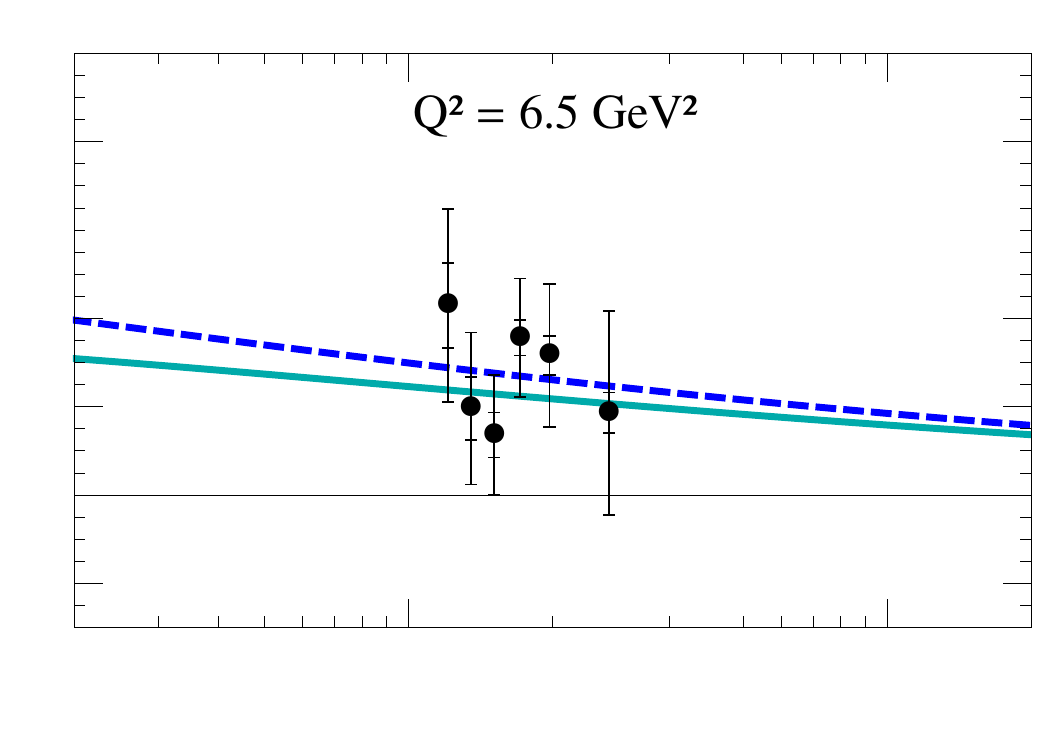}\hspace{-.17cm}
  \includegraphics[scale=.51]{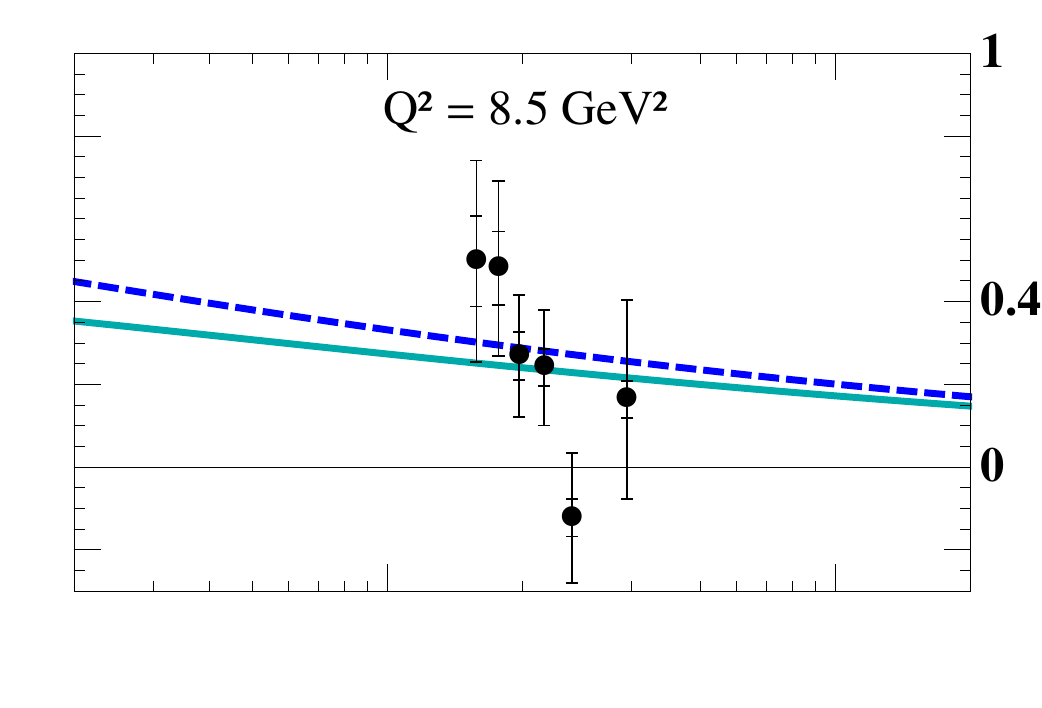}  
  \hspace{-2cm} \vspace{-.3cm}\\
  \hspace{-2cm}
  \includegraphics[scale=.52]{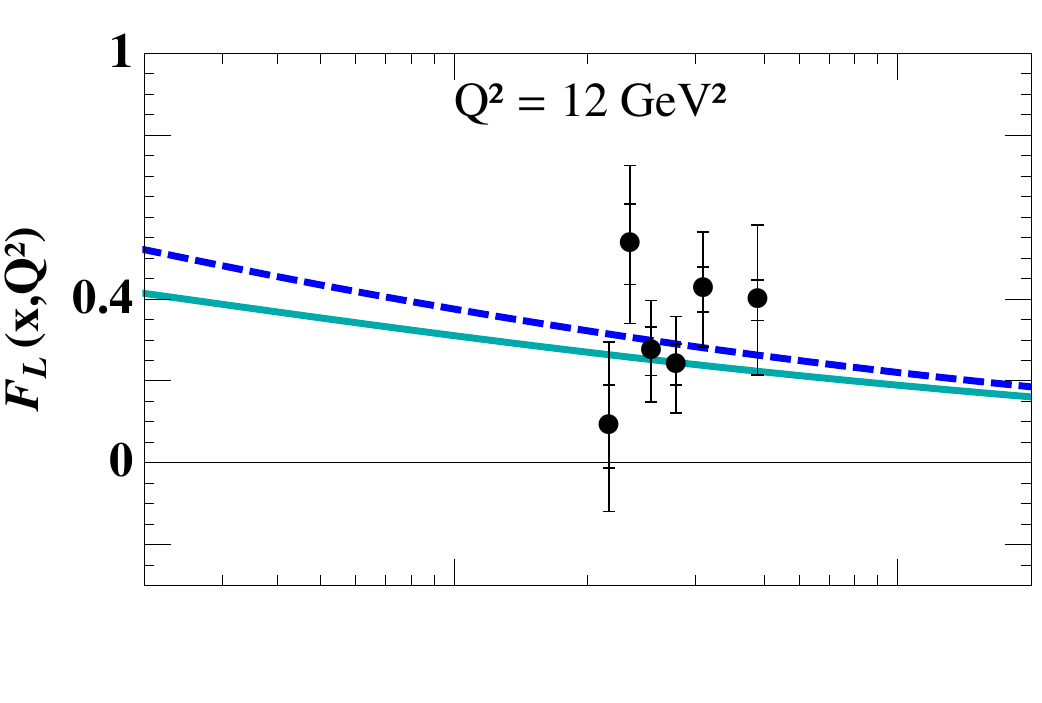} \hspace{-.33cm}
  \includegraphics[scale=.49]{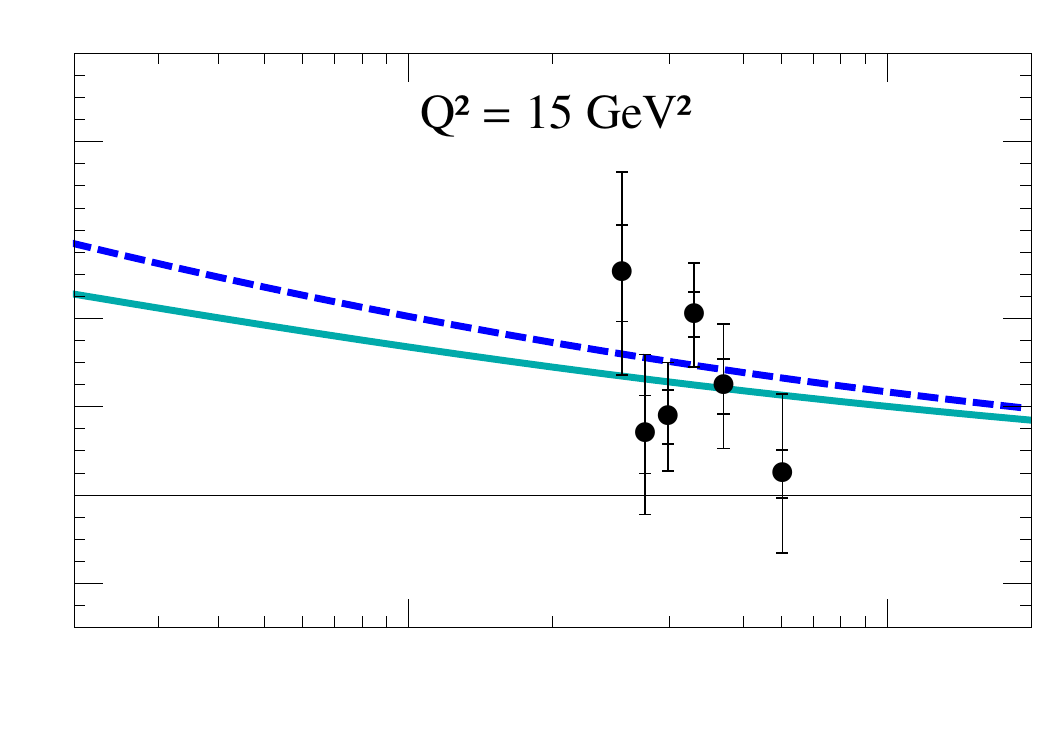}\hspace{-.17cm}
  \includegraphics[scale=.51]{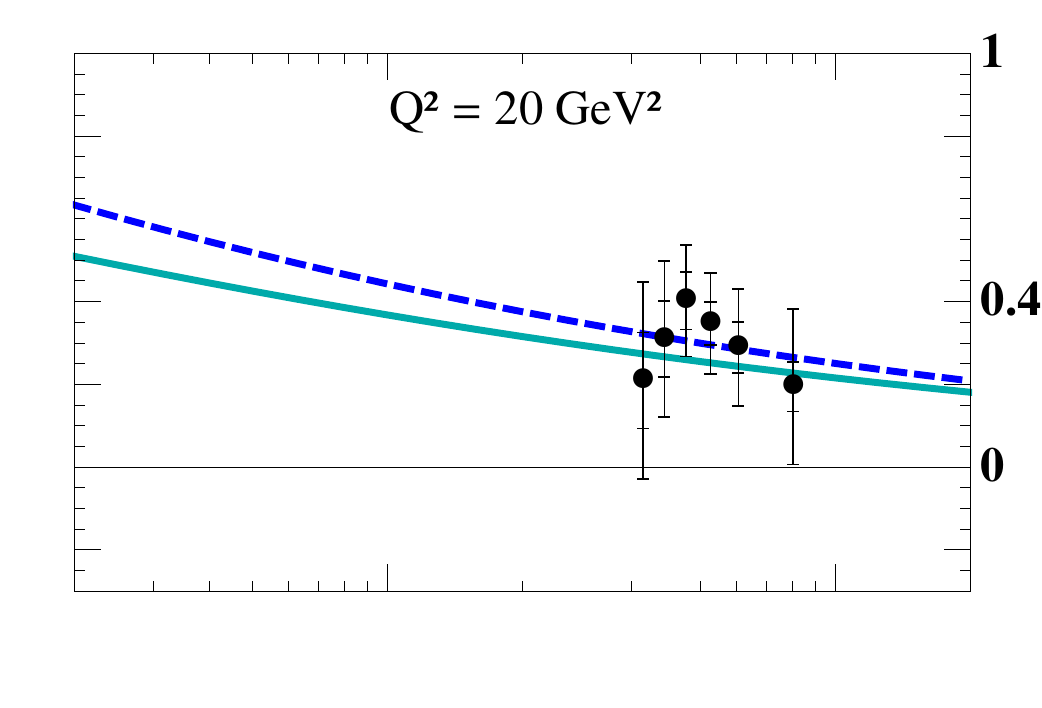} 
  \hspace{-2cm} \vspace{-.3cm} \\
  \hspace{-2cm}
   \includegraphics[scale=.52]{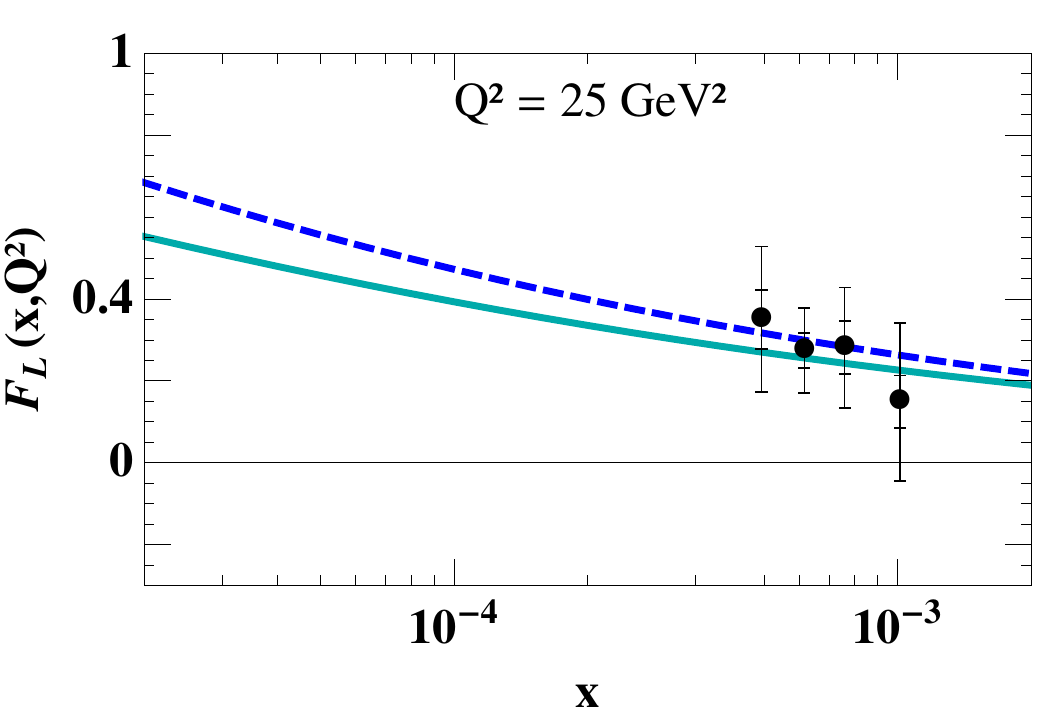} \hspace{-.33cm}
  \includegraphics[scale=.49]{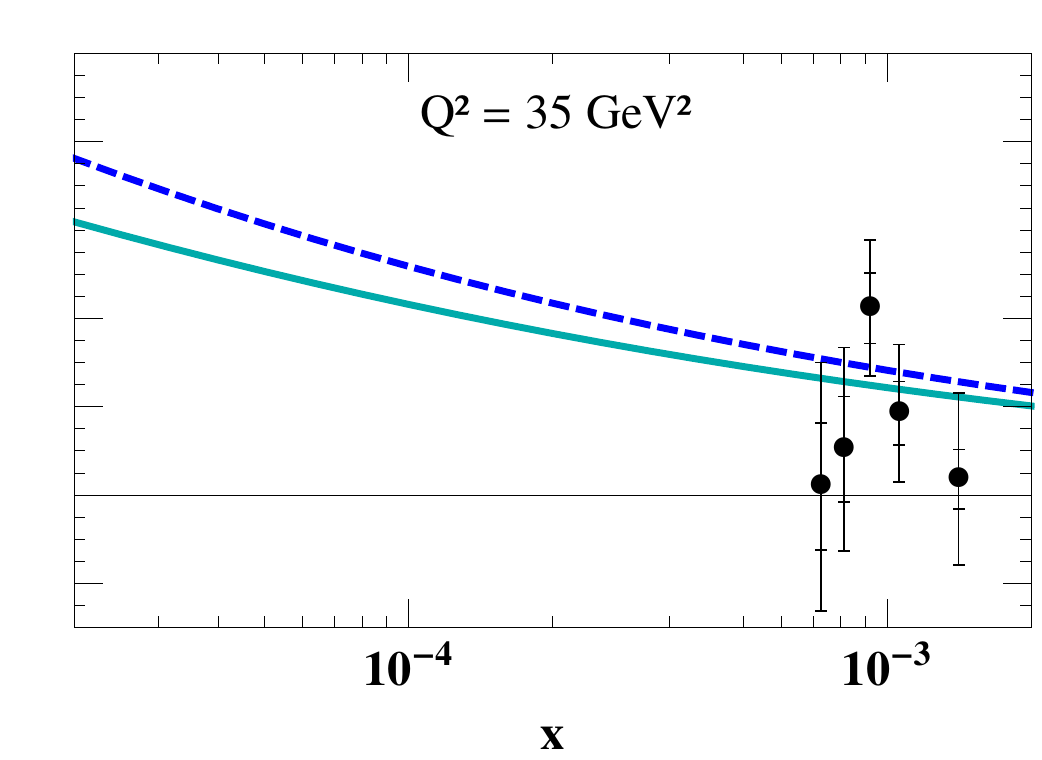}\hspace{-.17cm}
  \includegraphics[scale=.51]{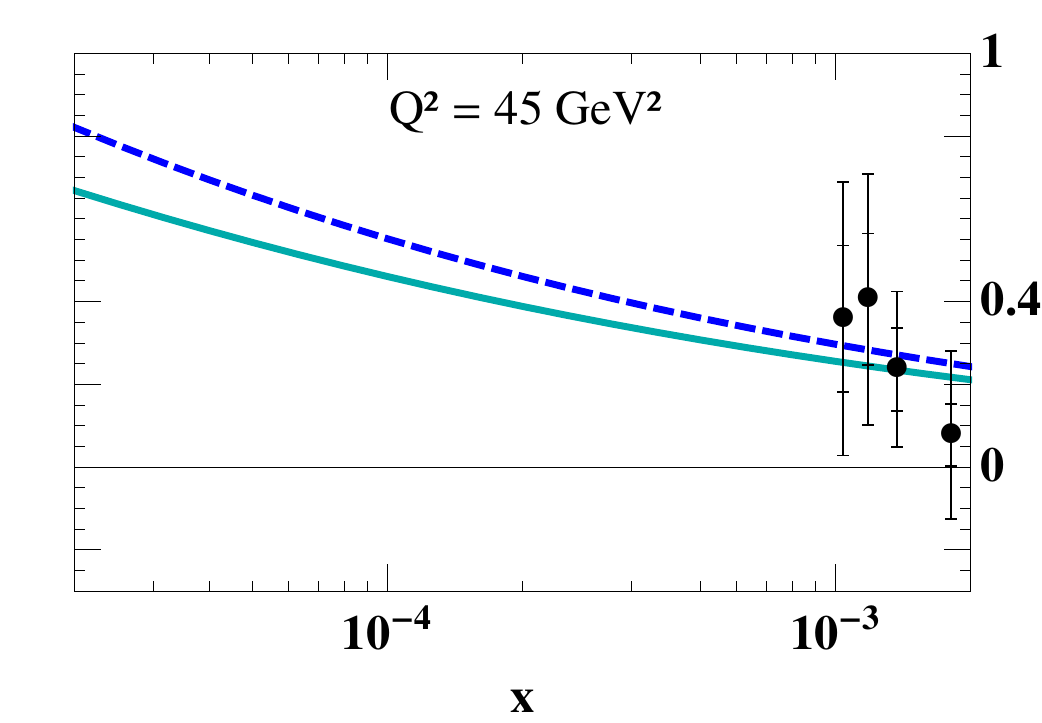} 
   \hspace{-2cm}
  \caption{Fit to $F_L$ with the LO photon impact factor (solid lines) and the improved one (dashed lines). The experimental data are taken from~\cite{Collaboration:2010ry}.}
  \label{fig:FL}
\end{figure}

\begin{figure}[htbp]
  \centering% \hspace{-3.5cm}
  \includegraphics[scale=.8]{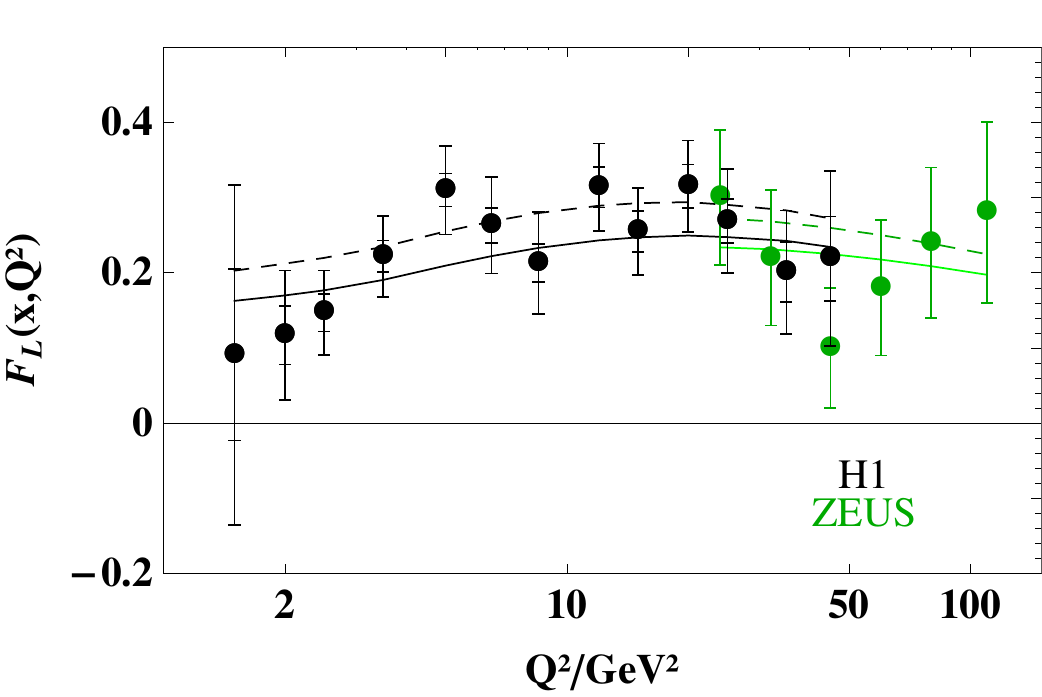}  
  \caption{The proton structure function $F_L$ as a function of $Q^2$. The average $x$ values for each $Q^2$ of the H1 data (black) are given in Figure 13 of~\cite{Collaboration:2010ry}. ZEUS data are taken from~\cite{Chekanov:2009na}. The solid line represents our calculation with the LO photon impact factor and the dashed line using the kinematically improved one.}
  \label{fig:FLvsQ2}
\end{figure}

The longitudinal structure function is an interesting observable which is very sensitive to the gluon content of the proton. 
We will now present our predictions for $F_L$ using the best values for the parameters previously obtained in the 
fit of $F_2$. We will see that the agreement with the data is very good. First, $Q^2$ is fixed and the $x$ dependence 
is investigated in Fig.~\ref{fig:FL}. The experimental data have been taken from~\cite{Collaboration:2010ry}. To 
present the $Q^2$ dependence it is convenient to calculate, for each bin in $Q^2$, the average value of $x$, see Fig.~\ref{fig:FLvsQ2}. In some 
sense this is a similar plot to the one previously presented for $\lambda$ in the $F_2$ analysis and we can see that the effect 
of using different types of impact factors is to generate a global shift in the normalization.  Again we note that 
we have an accurate description of the transition from high to low $Q^2$, which was one of the main targets of our work.


\section{Predictions for future colliders}

\begin{figure}[htbp]
  \centering \hspace{-1.7cm}
  \includegraphics[scale=.70]{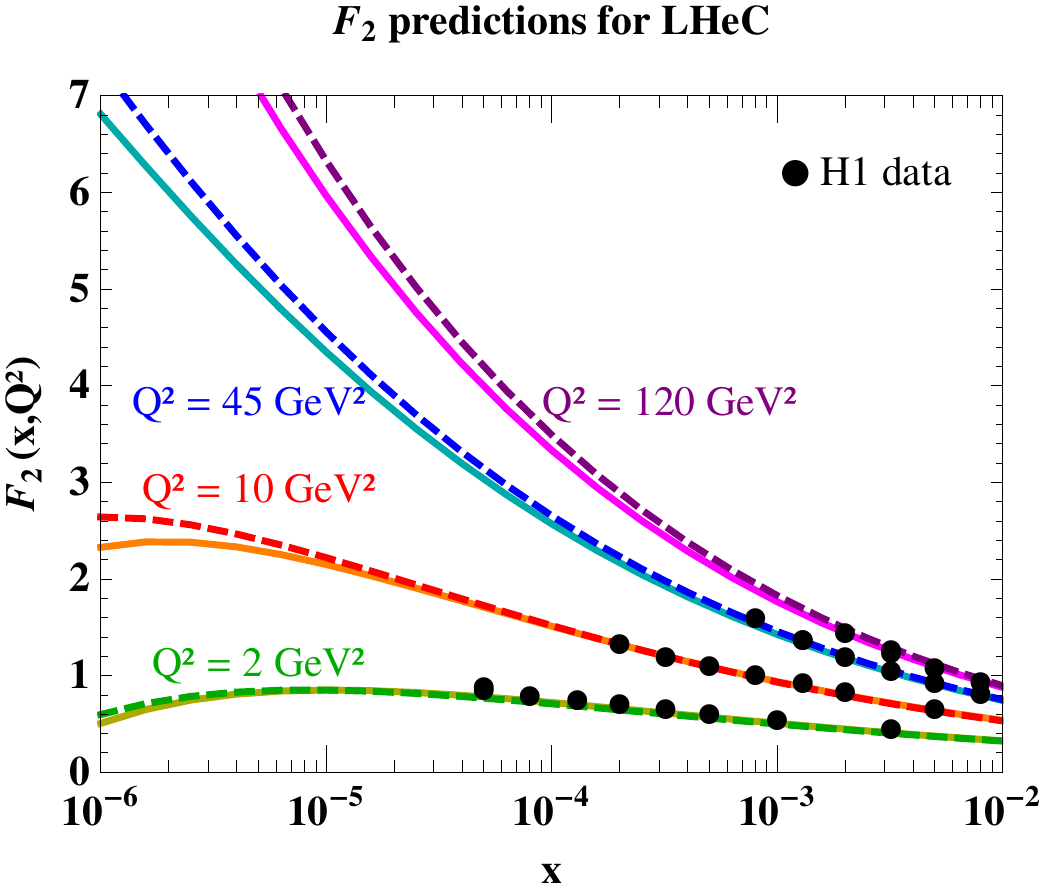}  
  \includegraphics[scale=.70]{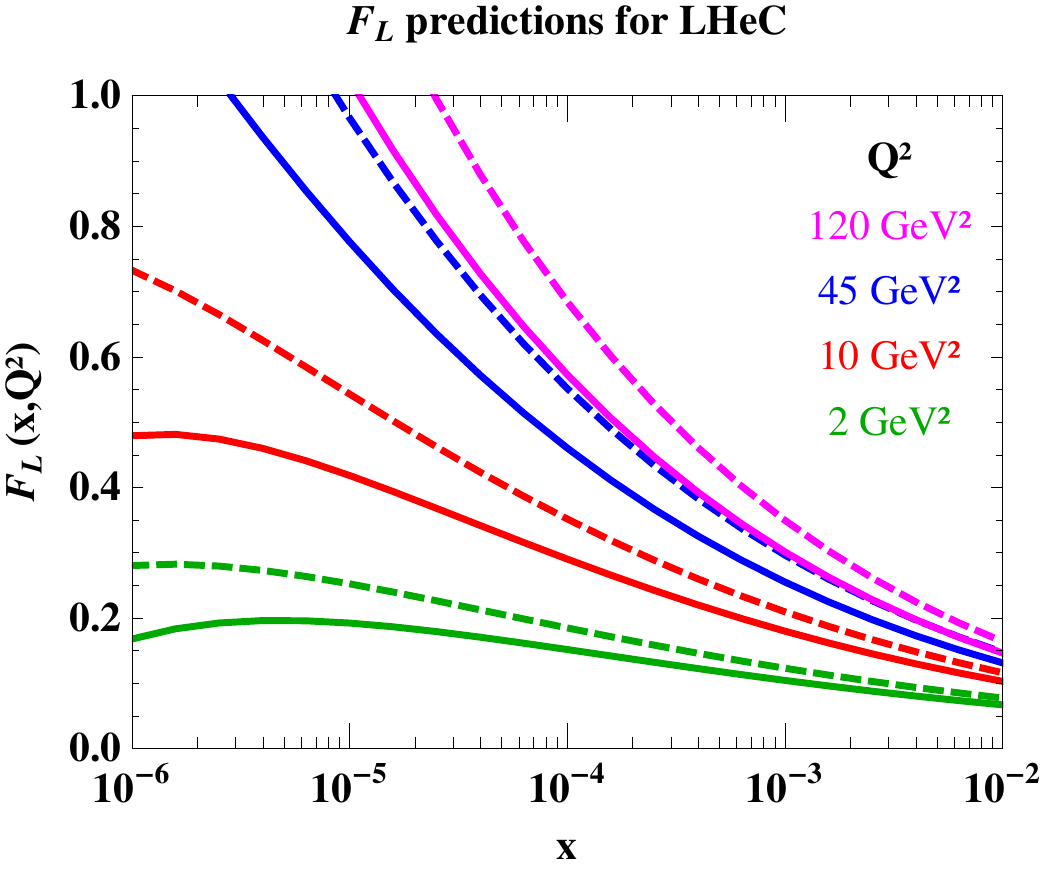}  
  \hspace{-1.7cm}
  \caption{Predictions for $F_2$ (left) and $F_L$ (right) for LHeC. On the left plot, the curve with $Q^2 = 10 \text{GeV}^2$ can be compared with Figure 4.13 of~\cite{AbelleiraFernandez:2012cc}. Simulated measurements for $F_L$ in the kinematic range plotted here (right) can be found in Figure 3.7 of the same reference.}
  \label{fig:FLvsQLHeC}
\end{figure}

While our predictions for the structure functions are in
agreement with the data from the HERA collider
experiments H1 and ZEUS, these observables are too inclusive to provide unambiguous evidence for BFKL evolution (for other recent studies in this context see~\cite{Kowalski:2010ue}). 
Comparable in quality fits can be obtained by both DGLAP evolution and
saturation models, see {\it e.g.} \cite{Collaboration:2010ry,
  Chekanov:2009na}. 
  % Even though qualitatively different, predictions from these formalisms agree within errors with the currently available data set. 
   In order to distinguish among different parton evolution pictures new collider experiments are needed, such as the
proposed Electron-Ion-Collider (EIC) at BNL/JLab (USA)
\cite{Deshpande:2012bu} and the Large Hadron Electron Collider (LHeC)
at CERN (Switzerland) \cite{AbelleiraFernandez:2012cc}, which will be
able to measure  both $F_2$ and $F_L$ at unprecedented small values of Bjorken $x$. 
In Fig.~\ref{fig:FLvsQLHeC} we present two studies with our predictions for 
$F_2$ and $F_L$ down to values of $x = 10^{-6}$. 

\section{Conclusions}

We have presented an application of the BFKL resummation program to the description of the $x$ and $Q^2$ 
dependence of structure functions as extracted from Deep Inelastic Scattering data at HERA. 
We have also provided some predictions for these observables at future colliders. 
In order to obtain the correct dependence on the 
virtuality of the photon at high values of the scattering energy, we have included in the BFKL kernel the main collinear contributions to all orders. We have also used optimal renormalization and an analytic running coupling in the infrared in order to accurately describe the regions of low $Q^2$. 
Our next task will be to use these parameterizations to describe more exclusive observables, such as heavy quark and multi-jet production, at the Large Hadron Collider at CERN.

\section*{Acknowledgments} 

We acknowledge partial support from the European Comission under
contract LHCPhenoNet (PITN-GA-2010-264564), the Comunidad de Madrid
through HEPHACOS S2009/ESP-1473, and MICINN (FPA2010-17747) and
Spanish MINECOs Centro de Excelencia Severo Ochoa Programme under
grant SEV-2012-0249.  M.H.  acknowledges support by the
U.S. Department of Energy under contract number DE-AC02-98CH10886 and
a BNL ``Laboratory Directed Research and Development'' grant (LDRD
12-034).

\end{document}